\documentclass[a4paper,onecolumn,aps,pra]{revtex4-2}
\usepackage[utf8]{inputenc}
\usepackage{amsmath} 
\usepackage{mathtools}
\usepackage{amsfonts} 
\usepackage{amssymb}
\usepackage{graphics}
\usepackage{graphicx}
\usepackage{float}
\usepackage{color}
\usepackage{braket}
\usepackage{tikz}
\usepackage{pgfplots}
\usepackage{bbold}
\usepackage{hyperref}
\usepackage{url}
\hypersetup{
    colorlinks=true,
    linkcolor=blue,
    citecolor=red,
    filecolor=magenta,      
    urlcolor=blue
    }
\usepackage{listings}
\usepackage{psfrag}
\usepackage{leftidx}
\usepackage{colordvi}
\usepackage{bbm}
\usepackage[caption=false]{subfig}

\newcommand{\dd}{\mathrm{d}}
\newcommand{\eE}{\mathrm{e}}

\newcommand{\Tr}{\mathrm{Tr}}
\newcommand{\nc}{\newcommand}
\nc{\ir}{\mathrm{i}}

\begin{document}

\title{Entanglement Hamiltonian for inhomogeneous free fermions}

\author{Riccarda Bonsignori}
\affiliation{%
  Institute of Theoretical and Computational Physics,
  Graz University of Technology, Petersgasse 16, 8010 Graz, Austria
}%

\author{Viktor Eisler}
\affiliation{%
  Institute of Theoretical and Computational Physics,
  Graz University of Technology, Petersgasse 16, 8010 Graz, Austria
}%

\begin{abstract}

We study the entanglement Hamiltonian for the ground state of one-dimensional free fermions in the presence of an inhomogeneous chemical potential. In particular, we consider a lattice with a linear, as well as a continuum system with a quadratic potential. It is shown that, for both models, conformal field theory predicts a Bisognano-Wichmann form for the entangement Hamiltonian of a half-infinite system. Furthermore, despite being nonrelativistic, this result is inherited by our models in the form of operators that commute exactly with the entanglement Hamiltonian. After appropriate rescaling, they also yield an excellent approximation of the entanglement spectra, which becomes asymptotically exact in the bulk of the trapped Fermi gas. For the gradient chain, however, the conformal result is recovered only after taking a proper continuum limit.

\end{abstract}

%\tableofcontents
%\newpage

\maketitle

\section{Introduction}

Entanglement properties provide invaluable insight to the characterization of ground states of quantum many-body systems \cite{AFOV-08,CCD-09,ECP-10,Laflorencie-16}. The information about how various parts of the system are coupled in the wavefunction is encoded in the reduced density matrix, obtained by integrating out all the degrees of freedom outside a given subsystem $A$. The analogy with the system-bath setup of statistical physics then suggests to write the reduced density matrix in an exponential form $\rho_A=\exp(-\mathcal{H})/Z$, where the entanglement Hamiltonian (EH) $\mathcal{H}$ has been the topic of numerous recent studies \cite{DEFV-23}. The most crucial observation is that, for a broad range of many-body ground states, the EH is well approximated by a \emph{local} operator, which has an intimate connection to the \emph{physical} Hamiltonian \cite{DEFV-23}. This finding is not only of high theoretical interest, but also allows to introduce efficient tomographic protocols \cite{Kokailetal21a}, such that the EH can be reconstructed in quantum-simulator experiments \cite{Joshietal23,Redonetal23}.

The emergence of such a simple local structure is best understood in the context of relativistic quantum field theory, where the EH is known as the modular Hamiltonian, and for a number of geometries can be written in the form
\begin{equation}
    \mathcal{H}= 2\pi \int_{A} \beta(x) T_{00}(x) \dd x \, .
    \label{eq:EHhom}
\end{equation}
Here $T_{00}(x)$ is the energy-density component of the stress tensor, while $\beta(x)$ is an appropriate weight function that depends on the subsystem $A$, and could also be interpreted as a local inverse temperature. In particular, if $A$ is the half of an infinite system, then the weight function is \emph{linear}, $\beta(x)=x$, which is the seminal result of Bisognano and Wichmann (BW) \cite{BW75,BW76}. While the BW result holds for quantum field theories in arbitrary dimensions, most of its generalizations require to consider a conformal field theory (CFT) in 1+1 dimensions. In particular, for an interval $A=[0,\ell]$ in an infinite line one finds a \emph{parabolic} weight $\beta(x)=x(\ell-x)/\ell$ \cite{HL82,CHM11}, and the results can further be extended to finite systems \cite{WKPZV-13,CT-16}, systems with boundaries or defects \cite{MT-21a,MT-21b}, and even to multiple interval geometries \cite{CH-09,LMR-10,ABCH-17}. In general, the inverse temperature $\beta(x)$ vanishes around the boundaries and increases towards the bulk of the subsystem, which clearly indicates that the entanglement is localized around the surface.

Although the result \eqref{eq:EHhom} is obtained for relativistic theories, a crucial question is how it translates to many-body systems, whose low-energy properties are captured by a CFT, but which break the Lorentz symmetry explicitly. The simplest example is a hopping chain of free fermions, where the analytic solution for the lattice EH of an interval shows deviations from the CFT result, and contains long-range hopping terms \cite{EP-17}. However, it was later shown that \eqref{eq:EHhom} can be exactly recovered via a continuum limit procedure \cite{ETP-19}, which can also be applied to different geometries \cite{ABCH-17,ETP-22} as well as for the EH of the massless harmonic chain \cite{DGT-20}. Moreover, it has been demonstrated for various critical quantum chains, that a simple discretization of the CFT ansatz \eqref{eq:EHhom} yields a very good approximation of the EH, reproducing the low-lying entanglement spectra and local observables to a high accuracy \cite{DVZ-18,GMCD-18,MGDR-19,ZCDR-20}. For continuum nonrelativistic critical systems, such as a Fermi gas with a quadratic dispersion, \eqref{eq:EHhom} must be interpreted as a deformation of the Schr\"odinger Hamiltonian, which gives again a perfect approximation of the EH \cite{E-23}.

In the context of free fermions, a further curiosity appears. Namely, for the hopping chain the discretized version of the CFT ansatz \eqref{eq:EHhom} corresponds to a \emph{tridiagonal} matrix, that commutes exactly with the lattice EH \cite{P-04}. In the continuum, the same role is played by a second order differential operator \cite{EP-13}. Remarkably, these commutation properties had been discovered long before in the context of time- and band-limited signals \cite{SP-61,Slepian-78,Slepian-83}, which translate directly to the free-fermion problem as the restrictions to the subsystem $A$ and the Fermi sea $F$ in the space and momentum domains, respectively. Commuting operators have also been identified for finite chains \cite{Grunbaum-81,EP-18}, and correspond again to the discretized CFT ansatz for the EH \cite{CT-16}. The existence of such commuting operators is believed to be rooted in the bispectrality property of the underlying eigenfunctions \cite{Grunbaum-94,GVZ-18,CNV-20}, and can even be extended to hopping chains with a particular inhomogeneity \cite{CNV-19,BCNPPV-22,BCNPV-24}.

Our goal here is to address the simplest possible inhomogeneous scenario for free fermions, namely the hopping chain with a linear, and the Fermi gas with a quadratic chemical potential. They both admit an effective low-energy description in terms of a Dirac fermion theory in curved space \cite{DSVC-17}, where the background metric is determined via the space-dependent Fermi velocity $v_F(x)$ induced by the inhomogeneity. Furthermore, the curved-space CFT framework even allows one to obtain the EH in the form \eqref{eq:EHhom}, where $\beta(x)$ depends on the metric \cite{TRLS-18}. However, we show that the interpretation of this weight in terms of an inverse temperature requires some care. Indeed, the latter should multiply the physical energy density, which depends on the local Fermi velocity. Hence, we rewrite $\beta(x)=\tilde\beta (x) v_F(x)$ and show that, in both of our examples, the inverse temperature $\tilde\beta (x)$ has a simple BW form, up to a rescaling with the Fermi velocity at the boundary. The above result applies to subsystems starting in the bulk of the fermionic density and extending to infinity, but finite-size effects can also be taken into account.

Furthermore, we show that the structure predicted by CFT is inherited by a tridiagonal matrix \cite{BOO-00} and a differential operator \cite{GRUNBAUM-83} that commute exactly with the EH in the lattice and continuum cases, respectively. Although they do not coincide with the exact EH, we demonstrate that they reproduce the low-lying entanglement spectra (and thus the entanglement entropy) to a very good accuracy, that improves when the length scale of the inhomogeneity is increased. Hence the BW ansatz provides an efficient local approximation of the EH, which does not suffer from numerical difficulties in obtaining its spectrum. In contrast, extracting the exact lattice EH requires very high-precision numerics, and leads to a non-local result with subdominant long-range hopping, such that the CFT form \eqref{eq:EHhom} can only be recovered after an appropriate continuum limit \cite{RSC-22}. Our scheme also allows to extend the results for the EH to nonrelativistic inhomogeneous free-fermion systems with more complicated potentials.

The rest of the paper is structured as follows. In Section \ref{sec:CFT} we discuss the CFT approach for calculating the EH of relativistic Dirac fermions with a space-dependent Fermi velocity. Our nonrelativistic inhomogeneous free-fermion models are introduced in Section \ref{sec:models} together with the methods to extract their EH. Namely, we consider a hopping chain with a linear potential, and a continuous Fermi gas with a quadratic potential, and our corresponding results are presented in Sections \ref{sec:gradient} and \ref{sec:trap}, respectively. The paper concludes with a discussion in Section \ref{sec:disc}, followed by three appendices containing technical details and derivation of some of the results.

\section{CFT approach}
\label{sec:CFT}

We start by summarizing the CFT technique that yields the expression of the EH for inhomogeneous systems \cite{TRLS-18}. To understand the main idea, we first discuss the procedure for a homogeneous chain of size $2L$ \cite{CT-16}, focusing on the simplest bipartition with the subsystem given by $A=[x_0,L]$. The reduced density matrix is then given by a path integral defined on the infinite strip $z \in [-L,L]\times \mathbb{R}$ in Euclidean spacetime, with a branch cut on the real axis ($\mathrm{Im}(z)=0$) along the subsystem $A$. It is useful to introduce a UV regularization around the entangling point by considering instead the segment $A_\epsilon=[x_0+\epsilon,L]$, such that the corresponding entanglement entropy is finite.

The goal is now to find a conformal map $z \to w$, which simplifies this geometry and transforms the strip into an annulus. This can be achieved by the transformation \cite{CT-16}
\begin{equation}
    w = f(z) = \ln \left[\frac{\sin\frac{\pi(z-x_0)}{4L}}{\cos\frac{\pi(z+x_0)}{4L}}\right],
    \label{eq:fz}
\end{equation}
such that $w \in [\ln R,0] \times [0,2\pi)$ lives in the logarithmic image of an annulus, where $\mathrm{Im}(w)=0$ is identified with $\mathrm{Im}(w)=2\pi$. Note that the inner radius $R \sim \epsilon$ of the annulus is finite due to the UV regularization, and the subsystem $A_\epsilon$ is mapped onto $f(A_\epsilon)=[\ln R,0]$. Therefore, the path integral in the transformed geometry simply corresponds to a thermal state of width $\ln R$ at inverse temperature $2\pi$, and the EH is given by
\begin{equation}
    \mathcal{H}= 2\pi \int_{f(A_\epsilon)} T_{00}(w) \dd w \, .
    \label{eq:EHw}
\end{equation}

To obtain the EH in the original geometry, one needs only the transformation rule of the energy-momentum tensor
\begin{equation}
    T(w) = \left(\frac{\dd z}{\dd w}\right)^2 T(z) +
    \frac{c}{12}\{z,w\},
    \label{eq:Tw}
\end{equation}
where $\{z,w\}$ is the Schwarzian derivative. Note that this term just gives a constant in \eqref{eq:EHw}, and can be absorbed into the normalization of the EH. Combined with the Jacobian of the transformation, one has in the original coordinates
\begin{equation}
    \mathcal{H}= 2\pi \int_{A} \frac{T_{00}(x)}{f'(x)} \, \dd x \, .
    \label{eq:EHz}
\end{equation}
The EH in a finite homogeneous system has then precisely the form \eqref{eq:EHhom}, with the weight function given by
\begin{equation}
\label{eq:betahom}
    \beta(x)=\frac{1}{f'(x)}=\frac{2L}{\pi}\frac{\sin\left(\frac{\pi x}{2 L} \right)-\sin\left(\frac{\pi x_0}{2 L} \right)}{\cos\left( \frac{\pi x_0}{2 L}\right)}.
\end{equation}

We now proceed to the CFT description of the inhomogeneous case, focusing on the massless free Dirac fermion with a space-dependent Fermi velocity $v_F(x)$. For simplicity, we consider only symmetric situations $v_F(-x)=v_F(x)$. As shown in Ref. \cite{DSVC-17}, this theory can be formulated as a Dirac fermion in \emph{curved space}, described by the action
\begin{equation}
    \mathcal{S}=\frac{1}{2\pi}\int \dd z \dd \Bar{z} \, \eE^{\sigma(x)}\left[ \psi_R^{\dagger} \tensor{\partial}_{\Bar{z}} \psi_R+ \psi_L^{\dagger} \tensor{\partial}_z \psi_L\right].
\end{equation}
The underlying Riemannian background metric
\begin{equation}
    \dd s^2=\eE^{2\sigma(x)}\dd z \dd \Bar{z}
    \label{eq:ds}
\end{equation}
is Weyl-equivalent to the flat metric, and the Weyl factor $\eE^{\sigma(x)}=v_F(x)$ plays the role of the local Fermi velocity.
It is then easy to see that the isothermal coordinates $z$ must be related to the original ones as
\begin{equation}
    z= \tilde{x}+ \ir t, \qquad  \tilde{x}=\int_0^x \frac{\dd y}{v_F(y)} \, .
    \label{eq:z}
%=\int_0^x  \eE^{-\sigma(y)}\dd y .
\end{equation}
Indeed, substituting these coordinates into the metric \eqref{eq:ds}, one obtains a simple local rescaling $\dd t \to v_F(x) \dd t$ of the time variable, which is exactly the expected effect of a space-dependent Fermi velocity. One can check that the same rescaling of time appears also in the local fermion propagator \cite{DSVC-17}.

The coordinate transformation \eqref{eq:z} changes also the domain of the path integrals. In particular, the half-width of the transformed strip becomes
\begin{equation}
\label{eq:tildeL}
    \tilde L = \int_0^L \frac{\dd y}{v_F(y)},
\end{equation}
and the new position of the entanglement cut $\tilde x_0$ follows analogously. However, the crucial observation of Ref. \cite{TRLS-18} is that the EH in the curved background has exactly the same form as the one defined on the flat metric $\dd z \dd \bar z$, parametrized by the new spatial coordinates $\tilde x$. In other words, one has
\begin{equation}
    \mathcal{H}= 2\pi \int_{\tilde A} \frac{T_{00}(\tilde x)}{\tilde f'(\tilde x)} \, \dd \tilde x \, ,
    \label{eq:EHinhom}
\end{equation}
where $\tilde A=[\tilde x_0,\tilde L]$, and the function $\tilde f$ is obtained from \eqref{eq:fz} by substituting the tilded coordinates $\tilde L$ and $\tilde x_0$. Finally, using again the property \eqref{eq:Tw}, one could rewrite \eqref{eq:EHinhom} in terms of the original coordinates $x$ as
\begin{equation}
    \mathcal{H}= 2\pi \int_{A} \frac{T_{00}(x)}{\tilde f'(\tilde x(x))\tilde x'(x)} \, \dd x \, = 2\pi \int_{A} \tilde \beta(x) v_F(x) T_{00}(x) \, \dd x \,,
    \label{eq:EHinhom2}
\end{equation}
where in the second step we used $\tilde x'(x)=v_F^{-1}(x)$, and the weight function reads explicitly
\begin{equation}    
\label{eq:betainhom}
    \tilde\beta(x)=\frac{2\tilde L}{\pi}\frac{\sin\left(\frac{\pi \tilde x(x)}{2 \tilde L} \right)-\sin\left(\frac{\pi \tilde x_0}{2 \tilde L} \right)}{\cos\left( \frac{\pi \tilde x_0}{2 \tilde{L}}\right)}.
\end{equation}

Although the result \eqref{eq:EHinhom2} is the same as in \cite{TRLS-18}, there is one crucial difference in its interpretation. Namely, in \cite{TRLS-18} the EH was written in the form \eqref{eq:EHhom}, i.e. the Fermi velocity was absorbed into the weight function $\beta(x)=\tilde \beta(x)v_F(x)$. However, in the actual models introduced in the next section, one would like to write the EH as a deformation of the physical energy density $H(x)$.
In the inhomogeneous setting, the local energy scale is fixed by $v_F(x)$, and thus $H(x)=v_F(x)T_{00}(x)$. Hence one expects that the deformation is described by the function \eqref{eq:betainhom}, which then has the correct interpretation of a spatially varying inverse temperature.

\section{Models and methods}
\label{sec:models}

In the following we introduce two different models, one defined on the lattice and the other in the continuum, whose low-energy properties are well described by a Dirac fermion in curved space.

\subsection{Gradient chain\label{subsec:gradient}}

We first consider a hopping chain with a linear potential
\begin{equation}
\label{eq:Hgrad}
    \hat H=-\frac{1}{2}\sum_{n=-L+1}^{L-1}(c_n^{\dagger}c_{n+1}+c_{n+1}^{\dagger}c_n)+\frac{1}{\xi}\sum_{n=-L+1}^{L}\left(n-\frac{1}{2}\right)c_n^{\dagger}c_n,
\end{equation}
where $c_n$ are fermionic annihilation operators at site $n$,
and the potential gradient is charactetized by the length scale $\xi$. The zero of the potential is chosen such that it is odd under a reflection $n\to 1-n$ of the chain. The single-particle eigenstates $\Phi_\kappa(n)$ of the Hamiltonian are given by a linear combination of Bessel functions $J_{n-\kappa}(\xi)$ and $Y_{n-\kappa}(\xi)$, where $\kappa$ depends on the boundary conditions \cite{Smith-71,Saitoh-73}. In our numerical calculations, we simply obtain $\Phi_\kappa(n)$ and the corresponding energies $\omega_\kappa$
by numerical diagonalization of the hopping matrix.

The ground-state of the chain is fully characterized by the correlation matrix
\begin{equation}
    C_{m,n}=\langle c_m^{\dagger}c_n\rangle=\sum_{\kappa \in F} \Phi_\kappa(m)\Phi_\kappa(n),
    \label{eq:Cgrad}
\end{equation}
where the sum runs over the Fermi sea $F$ of the occupied states,
satisfying $\omega_\kappa < 0$. The situation becomes particularly simple in the limit $L\to\infty$, where $\kappa$ must be an integer and the solutions read
\begin{equation}
    \Phi_\kappa(n)=J_{n-\kappa}(\xi), \qquad
    \omega_k =\xi^{-1}(\kappa-1/2).
\end{equation}
The spectrum is the famous Wannier-Stark ladder with equidistant energy levels \cite{Wannier-60}, and $F$ contains the states with $\kappa \le 0$. The sum in the correlation matrix \eqref{eq:Cgrad} can then be carried out explicitly and yields \cite{EIP-09}
\begin{equation}
    C_{m,n}=\frac{\xi}{2(m-n)}[J_{m-1}(\xi)J_n(\xi)-J_m(\xi)J_{n-1}(\xi)].
    \label{eq:Cgradinf}
\end{equation}
Note that this expression is unitarily equivalent to the one found for the domain-wall melting problem \cite{SK-23}, thus the two problems are intimately related. In particular, the gradient creates an interface with a non-trivial density profile of half-length $\xi$ between the completely filled/empty regions of the chain \cite{EIP-09}.

The entanglement properties of a subsystem $A$ can be determined by considering the eigenvalue problem of the reduced correlation matrix
\begin{equation}
\label{eq:RCM}
    \sum_{j\in A} C_{i,j} \, \phi_k(j) = \zeta_k \, \phi_k(i),
\end{equation}
where $i,j \in A$. The EH is a free-fermion operator \cite{Peschel03,PE-09}
\begin{equation}
  \label{eq:EHff} 
    \mathcal{H}= \sum_{i,j} H_{i,j}c_i^{\dagger}c_j,
\end{equation}
with the hopping matrix given by
\begin{equation}
    H_{i,j} =\sum_k \varepsilon_k \, \phi_k(i)\phi_k(j), \qquad
    \varepsilon_k = \ln \frac{1-\zeta_k}{\zeta_k}.
\end{equation}

To establish the connection with the curved-space CFT formulation of the previous section, we also need the local Fermi velocity $v_F(x)$. This can be obtained by applying a local density approximation (LDA), assuming a smoothly varying density profile ($\xi \gg 1$). The Hamiltonian is then considered to be homogeneous
around $x$, with an effective dispersion
\begin{equation}
    \omega_q(x) = -\cos q +x/\xi,
\end{equation}
such that the Fermi points are given by $|q_F|=\arccos(x/\xi)$.
The Fermi velocity then reads
\begin{equation}
    v_F(x) = \left.\frac{\dd \omega_q(x)}{\dd q}\right|_{q_F}=
    \sqrt{1-\left(\frac{x}{\xi}\right)^2}.
    \label{eq:vFgrad}
\end{equation}

\subsection{Fermi gas in a harmonic trap}

As a second example, we shall consider a continuous variable system, namely a trapped Fermi gas in a quadratic potential. The single-particle Hamiltonian for the harmonic trap is given by
\begin{equation}
    \hat H=-\frac{\hbar^2}{2m}\frac{\dd^2}{\dd x^2}+\frac{1}{2}m \, \Omega^2 x^2 - \mu,
    \label{eq:Htrap}
\end{equation}
where the chemical potential $\mu$ sets the particle number in the trap. The Hamiltonian can be made dimensionless by measuring length and energy in units of $\sqrt{\frac{\hbar}{m\Omega}}$ and $\hbar \Omega$, respectively, we thus set $\hbar=m=\Omega=1$ from here on. The oscillator wavefunctions and energy levels are then given by
\begin{equation}
\Phi_n(x)= c_n^{-1/2}
H_{n}(x) \eE^{-\frac{x^2}{2}}, \qquad
E_n=n+1/2,
\label{eq:Phitrap}
\end{equation}
where $H_n(x)$ with $n=0,1,\dots$ are the Hermite polynomials and the normalization factor reads $c_n=\sqrt{\pi}2^{n} n!$. The energy levels are thus again equidistant, and the $N$-particle ground state can be obtained by setting $\mu=N$. The correlation function reads
\begin{equation}
\label{eq:Kxy}
    K(x,y)=\sum_{n=0}^{N-1} \Phi_n(x)\Phi_n(y) .
\end{equation}

The notation $K(x,y)$ reflects that we now have an integral kernel
instead of a matrix. The spectrum then follows from the eigenvalue equation of the corresponding integral operator
\begin{equation}
    \label{eq:Kint}
    \mathcal{\hat K} \phi_k(x)=\int_{A} \dd y \, K(x,y) \phi_k(y) = \zeta_k \, \phi_k(x) .
\end{equation}
Alternatively, one could introduce the overlap matrix
\begin{equation}
\label{eq:overlapHar}
   \mathbb{A}_{m,n}= \int_A \dd x \, \Phi_m(x)\Phi_n(x),
\end{equation}
where the indices $m,n=0,\dots, N-1$ correspond to the occupied states. The matrix $\mathbb{A}$ then has exactly the same nontrivial ($\zeta_k \ne 0$) spectrum as $\mathcal{\hat K}$ \cite{CMV-11,CMV2-11}, however, their eigenfunctions are different. The operator form of the entanglement
Hamiltonian in the first-quantized notation is thus
\begin{equation}
    \mathcal{\hat H} = \ln (\mathcal{\hat K}^{-1}-1).
    \label{eq:EHtrap}
\end{equation}

Finally, we discuss the LDA for the trap. The effective dispersion of \eqref{eq:Htrap} reads
\begin{equation}
\omega_q(x) = \frac{q^2+x^2}{2} - N,
\end{equation}
where we have substituted $\mu=N$. This yields $|q_F|=\sqrt{2N-x^2}$ for the Fermi points, such that the density is given by the Wigner-semicircle with support $|x|<\sqrt{2N}$. Introducing the (dimensionless) half-length $\xi=\sqrt{2N}$ analogously to the gradient chain problem, one arrives at the Fermi velocity
\begin{equation}
    v_F(x) = \xi \sqrt{1-\left(\frac{x}{\xi}\right)^2}.
    \label{eq:vFtrap}
\end{equation}
The expression is almost identical to the one for the gradient chain \eqref{eq:vFgrad}, however, one obtains an extra factor $\xi$ and thus the maximal Fermi velocity scales with the number of particles in the trap.

\section{Entanglement Hamiltonian for the gradient chain\label{sec:gradient}}

Our first goal is to check how the CFT results presented in Section \ref{sec:CFT} can be applied to the simplest inhomogeneous lattice problem, the gradient chain. In general, the presence of the lattice yields a more complicated, not strictly local structure for the EH,
as observed already for the homogeneous hopping chain \cite{EP-17}. Nevertheless, the long-range hopping terms can be absorbed by considering a proper continuum limit, which reproduces the CFT result \eqref{eq:EHhom} with the proper weight function $\beta(x)$ \cite{ETP-19}. The procedure can also be generalized to inhomogenoeus settings, as demonstrated for the domain-wall melting problem \cite{RSC-22}, which is unitarily equivalent to the gradient chain.

In contrast, here we focus on the question, whether there exists a simple \emph{local} approximation of the EH, which reproduces the relevant part of the entanglement spectrum to a high accuracy. In other words, we try to find the discretized version of the CFT 
prediction in \eqref{eq:EHinhom2}. To get the inverse temperature profile $\tilde{\beta}(x)$, we first need to find the transformation $\tilde x(x)$ to curved-space coordinates in \eqref{eq:z}.
Using the expression of the Fermi velocity $v_F(x)$ in \eqref{eq:vFgrad}, we obtain
\begin{equation}
    \label{eq:xtilde}
    \tilde{x}(x)=\xi \arcsin\frac{x}{\xi},
    %\hspace{1 cm} \tilde{L}= \xi \arcsin \frac{L}{\xi}.
\end{equation}
whereas $\tilde L = \tilde x(L)$ and $\tilde x_0 = \tilde x(x_0)$. 
The expression of $\tilde{\beta}(x)$ then follows as
\begin{equation}
\label{eq:betagrad}
    \tilde{\beta}(x)=\frac{2}{\pi}\xi \arcsin\left( \frac{L}{\xi}\right)\frac{\sin \left( \frac{\pi}{2}\frac{\arcsin(x/\xi)}{\arcsin(L/\xi)}\right)-\sin \left( \frac{\pi}{2}\frac{\arcsin(x_0/\xi)}{\arcsin(L/\xi)}\right)}{\cos \left( \frac{\pi}{2}\frac{\arcsin(x_0/\xi)}{\arcsin(L/\xi)}\right)}.
\end{equation}
Clearly, the result depends on the ratios $L/\xi$ and $x_0/\xi$,
and we distinguish between the regimes of weak gradients ($L < \xi$) and the infinite-chain limit ($L \to \infty$), which will be treated separately.

\subsection{Infinite chain limit}
\label{subs:Inflatt}

Let us start by considering the simplest case of an infinite chain ($L\to \infty$), where the lattice problem simplifies considerably, as discussed in section \ref{subsec:gradient}. Note, however, that 
this regime corresponds to the ratio $L/\xi=1$ on the CFT side. Indeed, the coordinate transformation \eqref{eq:xtilde} is only defined in the regime $x\le \xi$ of nonvanishing Fermi velocity, as obtained from LDA. Considering larger chains $L>\xi$ should not affect the result, as the fermionic correlations decay exponentially
fast beyond that region. Setting $L=\xi$, the result \eqref{eq:betagrad} simplifies to
\begin{equation}
    \tilde \beta(x)=\frac{x-x_0}{\sqrt{1- \left(\frac{x_0}{\xi} \right)^2}},
\end{equation}
and thus the CFT prediction for the EH has exactly the BW form
\begin{equation}
\label{eq:EHLggxi}
    \mathcal{H}=\frac{2\pi}{v_F(x_0)} \int_{A}(x-x_0) H(x),
\end{equation}
rescaled by the local Fermi velocity at the entanglement cut.

The main question is now, how to recover this expression on the lattice. Let us choose a subsystem $A=[s+1,\infty)$, such that $s=0$ corresponds to the symmetric bipartition of the chain.
A crucial observation is that, analogously to the homogeneous case \cite{EP-17}, there exists a \emph{tridiagonal} matrix $T$ that commutes with the correlation matrix \eqref{eq:Cgrad} reduced to the subsystem, $[C_A,T]=0$. The commuting matrix reads \footnote{In \cite{BOO-00}, the authors consider the discrete Bessel kernel $\sum_{\kappa=1}^{\infty}J_{m+\kappa}(\xi)J_{n+\kappa}(\xi)$, where the summation index $\kappa$ starts from $1$ instead of $0$ as in \eqref{eq:Cgradinf}, such that their matrix corresponds to $C_{m+1,n+1}$ in our notation. However, their kernel is restricted to the subsystem $[s,\infty)$, which corresponds to the restriction of \eqref{eq:Cgradinf} to the indices $[s+1,\infty)$, as considered in sec.~\ref{subs:Inflatt}.}\cite{BOO-00}
    \begin{equation}
        T=\begin{pmatrix}
            d_1 & t_1 &  &  &  \\
            t_1 & d_2 & t_2 &  \\
             & t_2 & d_3 & t_2 \\
             &  & \ddots & \ddots & \ddots \\
        \end{pmatrix},
    \end{equation}
with matrix elements given by
\begin{equation}
\label{eq:Telements}
    t_i=i-s, \hspace{1.5 cm}  d_i=- \frac{2}{\xi}\left(i-\frac{1}{2}\right)\left(i-\frac{1}{2}-s\right).
\end{equation}
The $t_i$ can be interpreted as local hopping amplitudes, which increase linearly from the boundary $s$ of the subsystem. The diagonal chemical potentials $d_i$ follow a parabolic profile, and can be interpreted as the gradient term in the physical Hamiltonian \eqref{eq:Hgrad} multiplied by the BW weight factor, that is now discretized at half-integer values. One should stress that the diagonal terms are defined only up to an arbitrary constant, and the above choice was guided by the procedure outlined below.

Although the matrix $T$ has precisely the structure \eqref{eq:EHLggxi} dictated by CFT, and even commutes with the lattice EH, $[H,T]=0$, their spectra are different. The situation is thus very similar to the one observed for the homogeneous chain \cite{EP-17}, which allows to write the EH as a power series
\begin{equation}
H=\sum_{r=0}^\infty \alpha_r T^r.
\label{eq:HT}
\end{equation}
While $\alpha_0$ and $\alpha_1$ simply correspond to a constant shift and a rescaling of the $T$ matrix, higher powers with $r>1$ modify the structure of $H$ and introduce hopping at distance $r$.
The coefficients $\alpha_r$ are, in principle, determined by matching the spectra $\varepsilon_k$ and $\lambda_k$ of the respective matrices $H$ and $T$. While in the homogeneous case this is possible using the uniform asymptotic expansions of both spectra \cite{Slepian-78}, for the gradient problem such results are not available, and we follow a different route. Indeed, it is reasonable to expect that the low-lying spectrum should be well reproduced by discarding all the $r>1$ terms in \eqref{eq:HT}, which yields the local approximation of the EH. Comparing the amplitudes \eqref{eq:Telements} to the physical energy density in \eqref{eq:Hgrad}, the CFT result \eqref{eq:EHLggxi} suggests the scale factor $\alpha_1=-\pi/v_F(s)$.

%%%%%%%%%%%%%%%%%%%%%%%%%%%%%%%%%%%%%%%%%%%%
\begin{figure}[t]
\centering
    \includegraphics[width=0.55\textwidth]{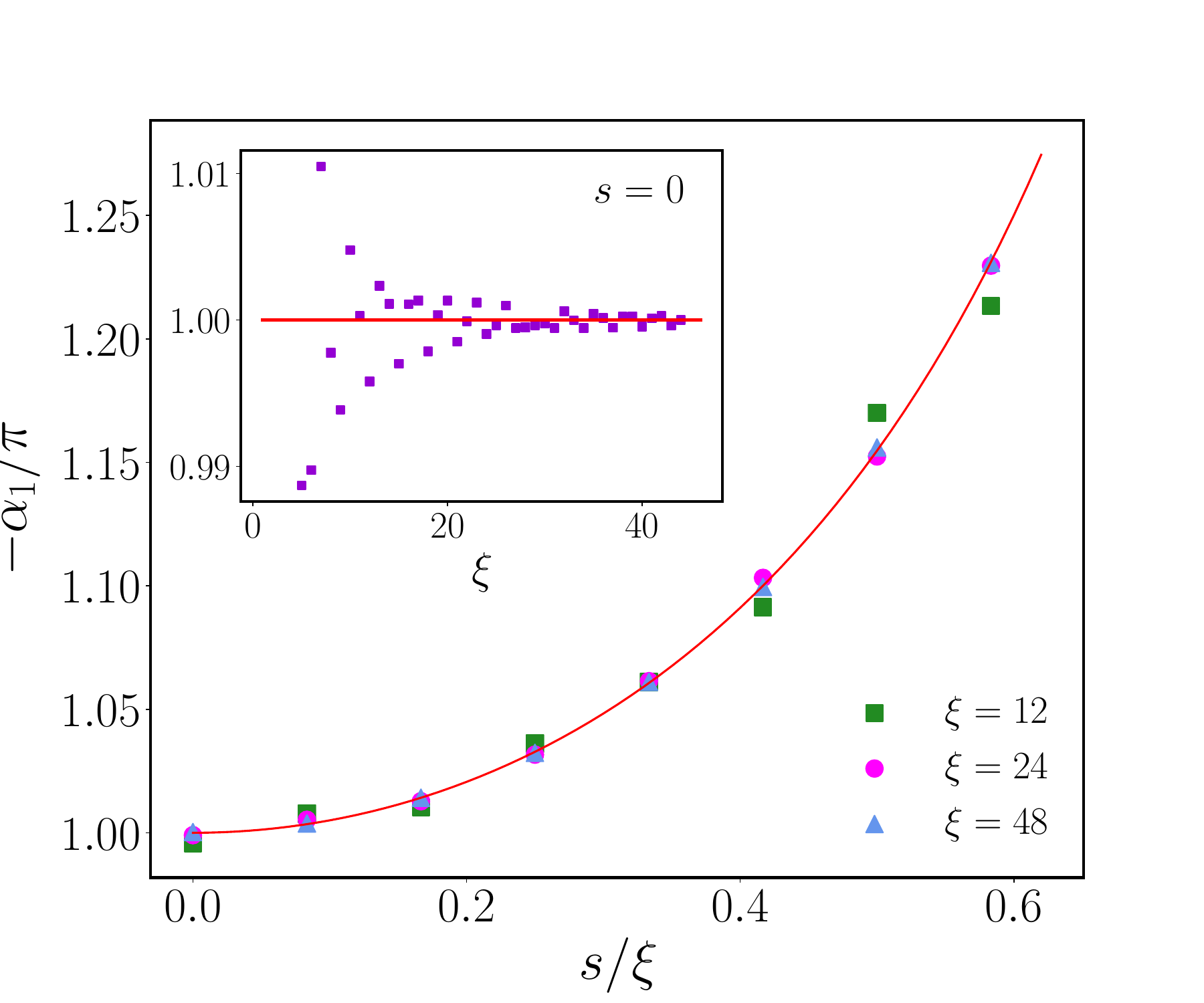}
      \caption{Scaling factor $-\alpha_1/\pi$ (symbols) obtained from the minimization procedure described in the text, as a function of $s$ and various $\xi$. The BW prediction $1/v_F(s)$ is shown by the red solid line. The inset shows the convergence of the same factor at fixed $s=0$ and increasing $\xi$.}
    \label{fig:spectraLgxi}
\end{figure}
%%%%%%%%%%%%%%%%%%%%%%%%%%%%%%%%%%%%%%%%%%%%

In order to have an unbiased estimate, we used a fitting  procedure suggested in \cite{BMPV-23}. To this end, let us introduce the local BW-ansatz for the density matrix
\begin{equation}
    \rho_{BW}=\frac{\eE^{-\mathcal{H}_{BW}}}{Z_{BW}}, \qquad
    \mathcal{H}_{BW}=\sum_{i,j\in A}(\alpha_0\delta_{i,j}+ \alpha_1 T_{i,j})c_i^{\dagger}c_j .
    \label{eq:HBW}
\end{equation}
The main idea is to find the best match for the low-lying single-particle spectra of $\rho_A$ and $\rho_{BW}$, by comparing expectation values of operators that suppress the contributions from the high-energy spectrum. The simplest choice is to use the entanglement entropy
\begin{equation}
\label{eq:S}
    S=-\Tr (\rho_A \log \rho_A)=
    \sum_{k} \frac{\varepsilon_k}{\eE^{\varepsilon_k}+1}+ \ln (1+\eE^{-\varepsilon_k}),
\end{equation}
while the corresponding BW-approximation follows by replacing $\varepsilon_k \to \alpha_0 + \alpha_1\lambda_k$. The best fit is then achieved by minimizing the deviation $\delta S=S-S_{BW}$. Since we included also the constant $\alpha_0$, which acts as a chemical potential, we need to minimize also the deviation $\Tr[(\rho-\rho_{BW}) \hat N_A]$ between the expectation values of the particle number $\hat N_A=\sum_{i\in A} c_i^{\dagger}c_i$. In fact, the result of this minimization for $\alpha_0$ guided us to choose the proper discretization for the $d_i$ in \eqref{eq:Telements}.
%%%%%%%%%%%%%%%%%%%%%%%%%%%%%%%%%%%%%%%%%
\begin{figure}[H]
    \hspace*{-0.2cm}
    %\scalebox{1}{\input{./PLOTS/TEX/1GS_ceff_kappa.tex}}\\
    \subfloat{\includegraphics[width=0.45\textwidth]{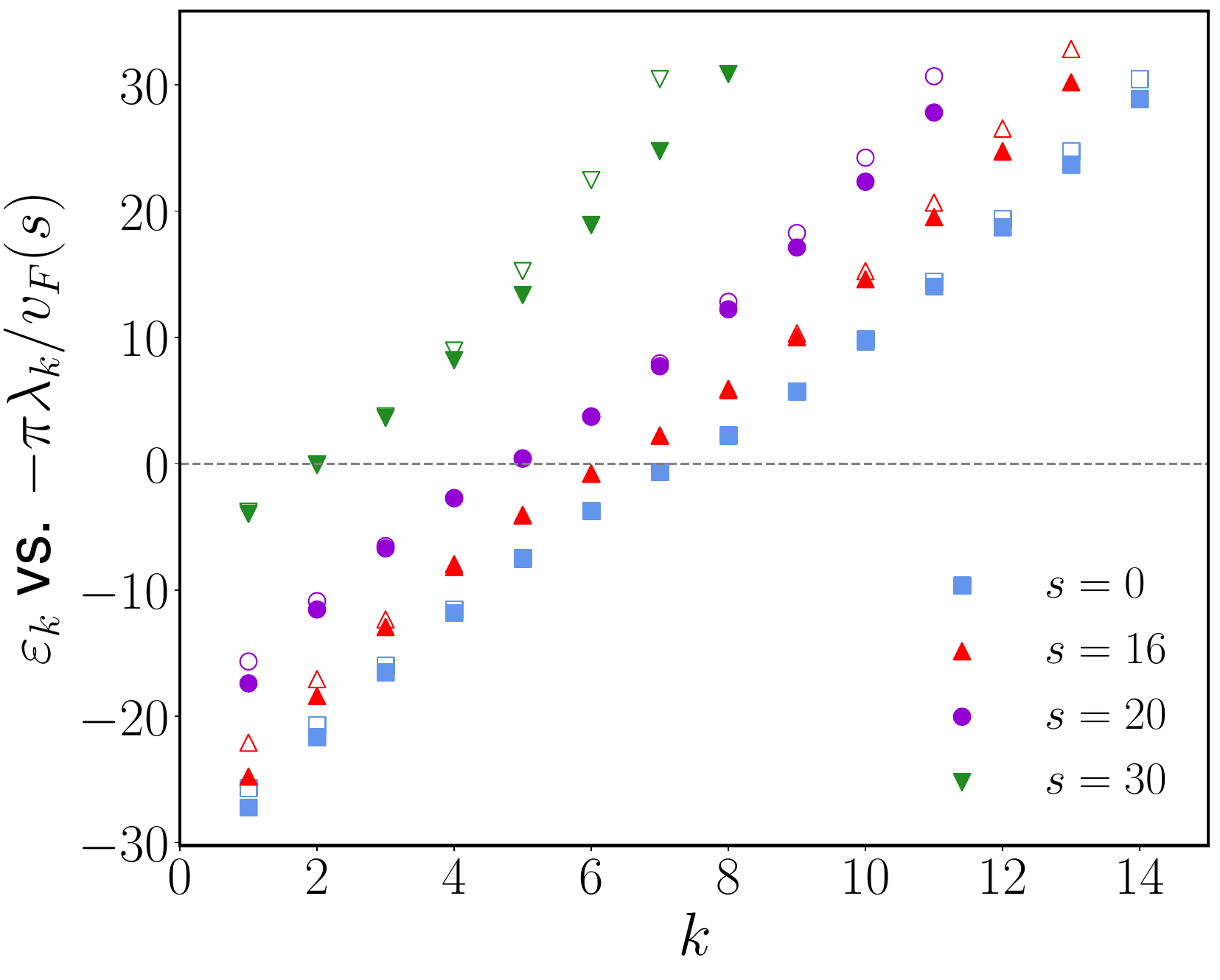}}
    \hspace{0.5cm}\subfloat{\includegraphics[width=0.45\textwidth]{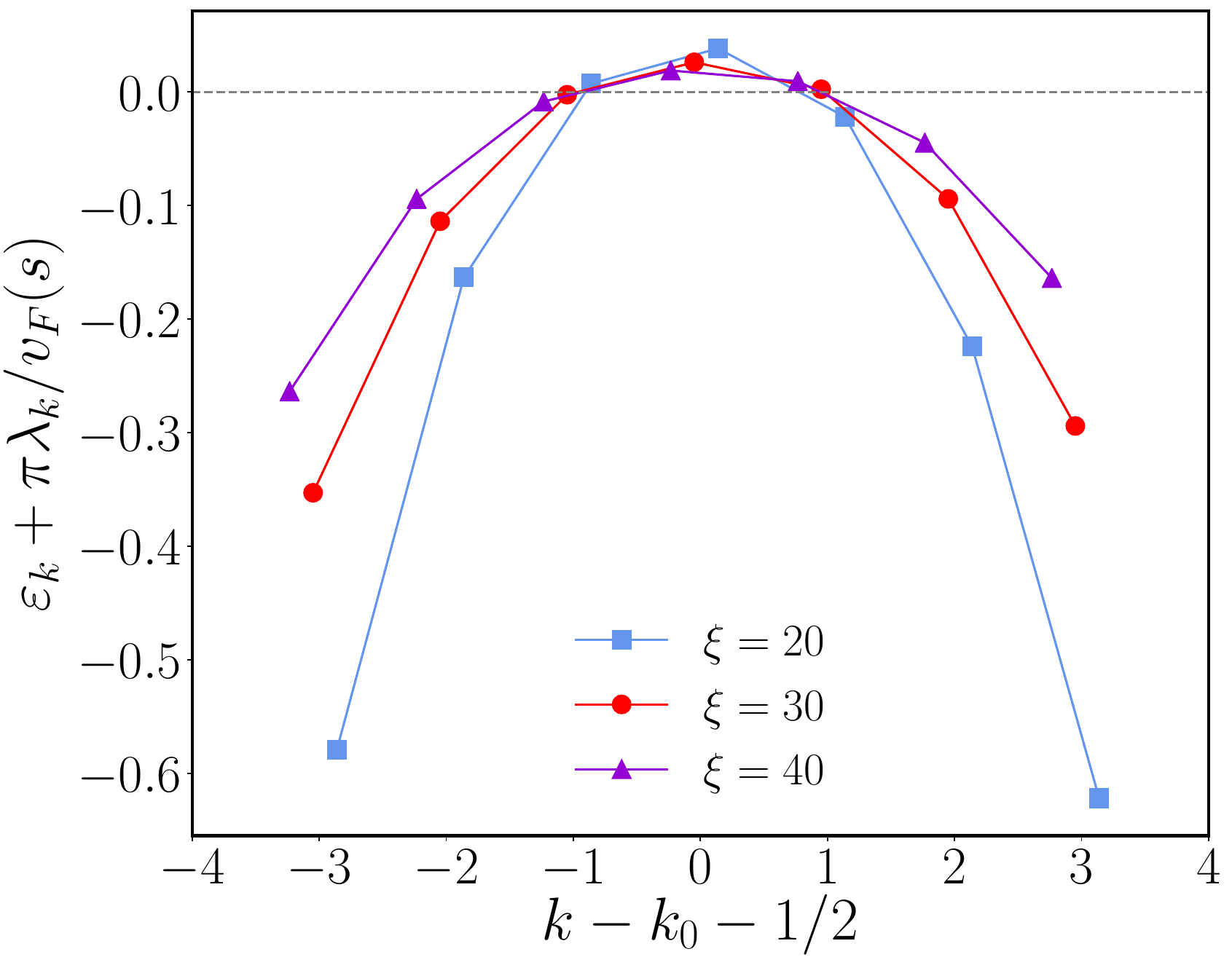}}
   \caption{\textit{Left:} single-particle eigenvalues $\varepsilon_k$ of $\mathcal{H}$ (full symbols) compared to those $-\pi \lambda_k /v_F(s)$ of $\mathcal{H}_{BW}$ (empty symbols) for the gradient chain with different $s$, $L=60$ and $\xi=40$. \textit{Right:} difference between the spectra for fixed $s=0$, and various $\xi$.}
     \label{fig:spectragrad}
\end{figure}
%%%%%%%%%%%%%%%%%%%%%%%%%%%%%%%%%%%%%%%%%

%%%%%%%%%%%%%%%%%%%%%%%%%%%%%%%%%%%%%%%%%%%%
\begin{figure}[H]
\centering    \includegraphics[width=0.5\textwidth]{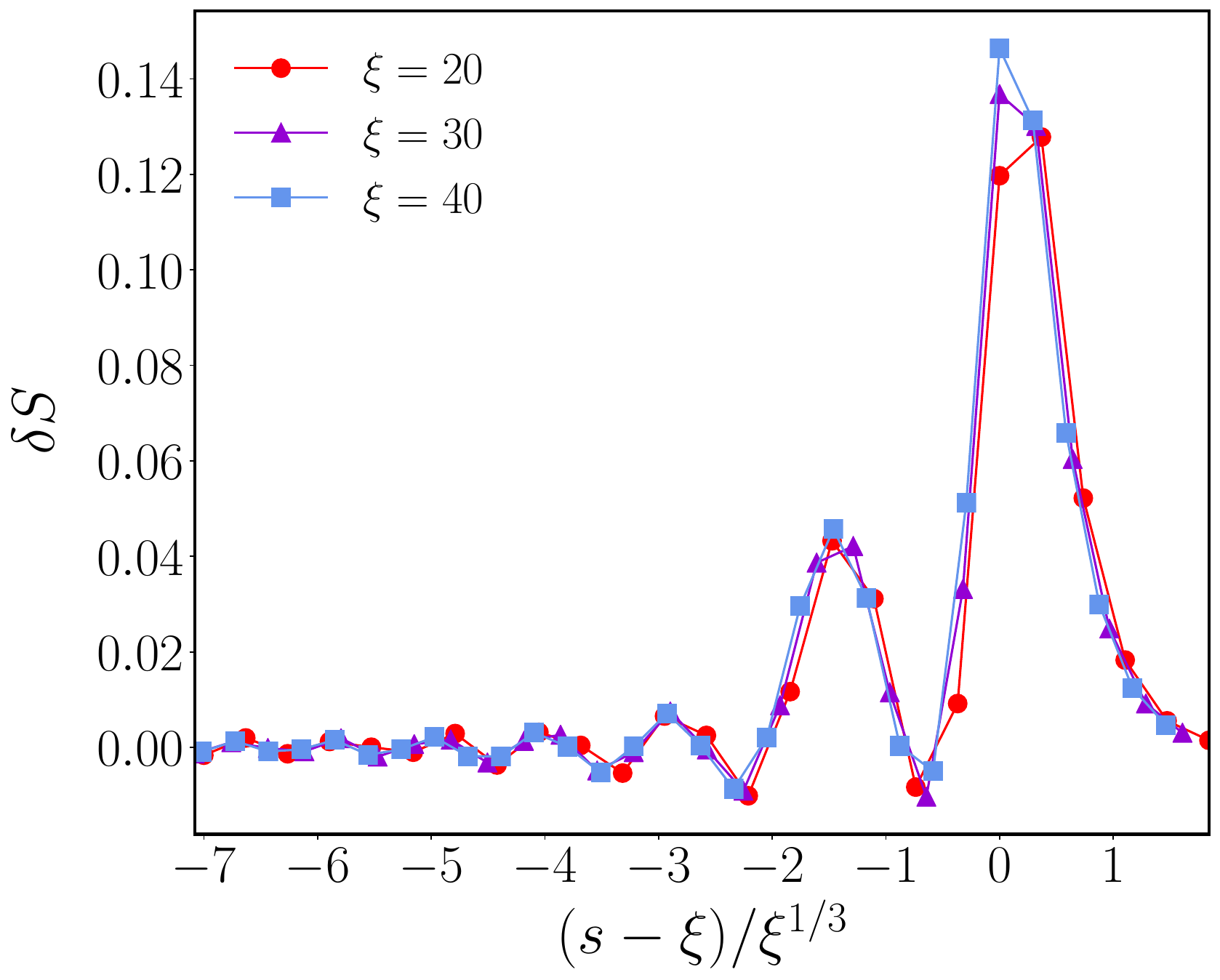}
      \caption{Entropy difference $\delta S=S-S_{BW}$ plotted against the edge scaling variable $(s-\xi)/\xi^{1/3}$ for increasing values of $\xi$.}
    \label{fig:deltaSedge}
\end{figure}
%%%%%%%%%%%%%%%%%%%%%%%%%%%%%%%%%%%%%%%%%%%%

 For the scale factor $\alpha_1$, the results of the minimization are shown in Fig. \ref{fig:spectraLgxi} as a function of the scaled subsystem boundary $s/\xi$.
 The symbols correspond to data sets with different values of $\xi$, and a good agreement is found with the analytical prediction $\alpha_1=-\pi/v_F(s)$, shown by the solid red line. The small deviations from this curve observed in the data, especially for larger values of $s$, diminish as $\xi$ increases. This is magnified in the inset for fixed $s=0$, where the fitted values approach the analytical prediction in an oscillatory manner for increasing $\xi$. One should note that the numerical data was obtained for a finite chain with fixed $L=60$, which is sufficiently larger than the interface length $\xi$. Due to the exponentially small matrix elements $C_{i,j}$ for $i,j>\xi$, a further increase of $L$ does not have any noticeable effect on the low-lying eigenvalues $\varepsilon_k$, and the same is true for $\lambda_k$. Thus the finiteness of $L$ does not influence the results, as long as $L \gg \xi$ is satisfied.

With the parameters of $\mathcal{H}_{BW}$ fixed, we now have a look at the corresponding spectra to assess the quality of the BW approximation. These are compared against the actual spectra, shown by the full and empty symbols, respectively, for a fixed $\xi$ and various $s$ on the left of Fig.~\ref{fig:spectragrad}. The match is indeed excellent for the low-lying eigenvalues, and deteriorates only around the bottom and top of the spectrum for small $s$. For increasing $s$, however, the discrepancies grow and move towards the center of the spectrum. For a fixed $s$, the mismatch between the spectra diminishes for increasing $\xi$, as illustrated on the right of Fig.~\ref{fig:spectragrad} for $s=0$. In order to align the spectra, a constant $k_0$ has been subtracted from the spectral index, which is known to be given by the expected particle number $\braket{\hat N_A}$ from earlier work \cite{EP-14}. The difference has then a roughly parabolic form which flattens around $k \approx k_0$ for increasing $\xi$. Note, however, that the maxima of the curves are slightly positive, implying the presence of finite-size corrections even within the low-lying region of the spectrum. This also explains the slight deviations of $\alpha_1$ from the analytical form in Fig.~\ref{fig:spectraLgxi}, resulting from the fitting procedure.

The results above show that the approximation $\varepsilon_k=-\pi\lambda_k/v_F(s)$, corresponding to the discretized BW-ansatz, works indeed very well if the boundary $s$ of the subsystem is still within the bulk of the interface. Instead, for $s \to \xi$ one approaches the edge regime, where a different kind of scaling behaviour is known to appear \cite{HRS-04,ER-13,EP-14,JMSthephan-19,Gouraud-24}. This is characterized by the Airy kernel \cite{TW-94} and a length scale $\xi^{1/3}$, such that the LDA-based CFT approach breaks down in the dilute edge regime. To illustrate this, we plot in Fig.~\ref{fig:deltaSedge} the entropy difference $\delta S$ against the edge scaling variable $(s-\xi)/\xi^{1/3}$ for increasing values of $\xi$. One observes a data collapse with a scaling function peaked around $s=\xi$, and having a quick oscillatory decay towards the bulk. This is a clear indication for the breakdown of the BW ansatz if the subsystem boundary is located within the edge regime.

\subsection{Small gradients}

We move on to consider cases where the length scale induced by the gradient is larger than the chain length, $\xi > L$. For simplicity, we focus on the case $s=0$. The structure of $C_A$ is then more complicated, and a commuting tridiagonal matrix is not known in the literature. One could, however, construct again a $T$ matrix as the discretized version of the CFT result \eqref{eq:betagrad}, with matrix elements given by
\begin{equation}
%    \begin{split}
        \label{eq:Ttilde}
        t_{i}= \tilde \beta(i), \qquad
        %\xi \frac{2}{\pi}\arcsin(R)\sin\left(\frac{\pi}{2} %\frac{\arcsin(i/\xi)}{\arcsin(R)}\right), \\
         d_{i}=-\frac{2}{\xi} \left(i-\frac{1}{2}\right)\tilde \beta(i-1/2).
         %\frac{2}{\pi}\arcsin(R)\sin\left(\frac{\pi}{2} \frac{\arcsin((i-1/2)/\xi)}{\arcsin(R)}\right),
%    \end{split}
\end{equation}
We set $x_0=0$ and fix the ratio $R=L/\xi$, such that in the limit $R\to 1$ the expressions reduce to \eqref{eq:Telements}.

%%%%%%%%%%%%%%%%%%%%%%%%%%%%%%%%%%%%%%%%%
\begin{figure}[t]
    \centering   
    \hspace*{-0.25cm}
    \subfloat{\includegraphics[width=0.45\textwidth]{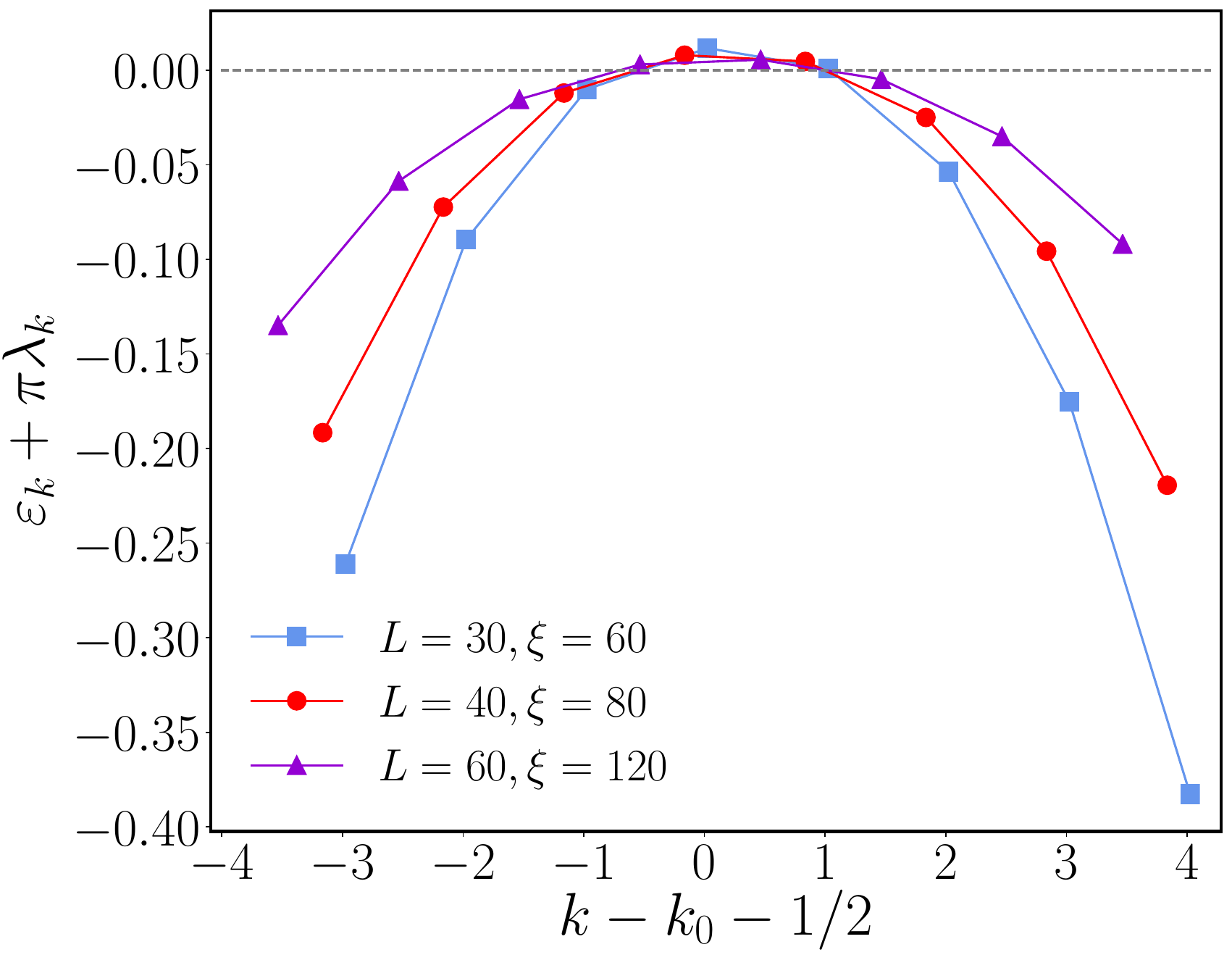}}
    \qquad
    \subfloat{\includegraphics[width=0.45\textwidth]{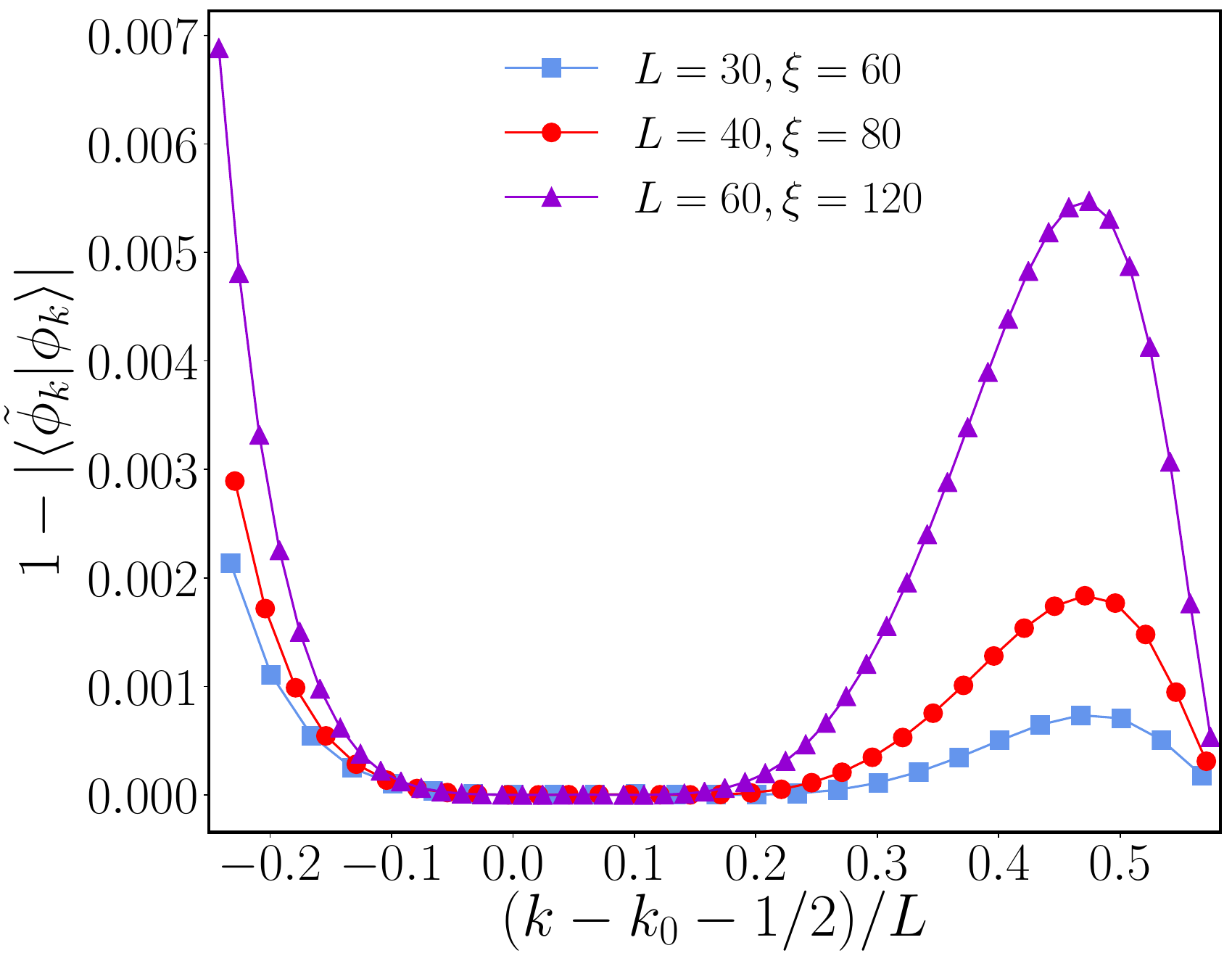}}
     \caption{\textit{Left:} deviations between the eigenvalues of $H$ and $-\pi T$, with the elements of the tridiagonal matrix $T$ defined in \eqref{eq:Ttilde}. The data are shown, with an appropriate shift, only for the low-lying part of the spectra, for a fixed ratio $R=1/2$ and increasing $L$.
     \textit{Right:} deviation of the corresponding eigenvectors $\phi_k$ and $\tilde{\phi}_k$, as measured by their overlap, shown against the scaled spectral index.
}
     \label{fig:smallgrad}
\end{figure}
%%%%%%%%%%%%%%%%%%%%%%%%%%%%%%%%%%%%%%%%%

%%%%%%%%%%%%%%%%%%%%%%%%%%%%%%%%%%%%%%%%%%%%
\begin{figure}[t]
\centering    \includegraphics[width=0.5\textwidth]{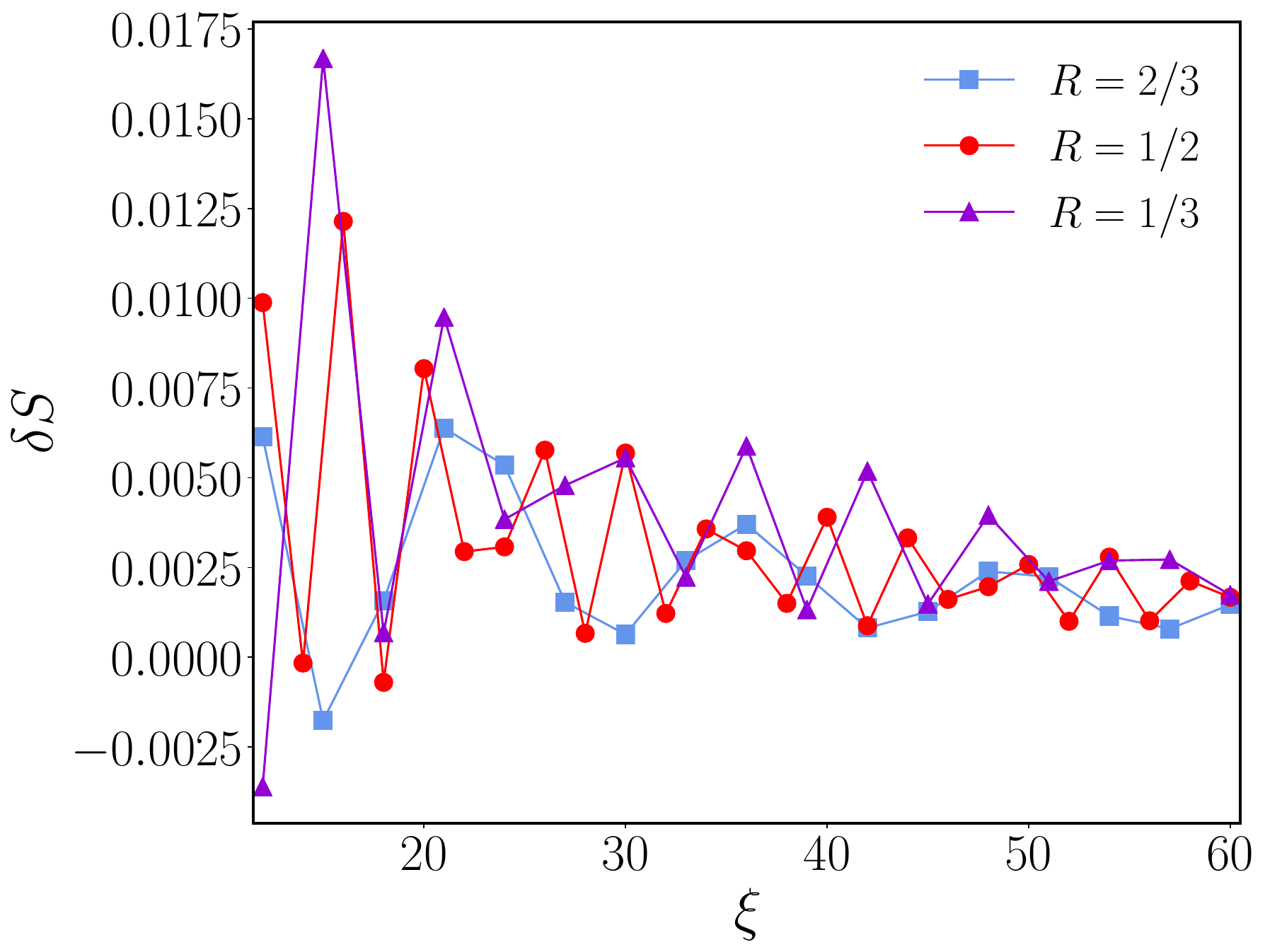}
      \caption{Difference of the entanglement entropy for various ratios $R=L/\xi$ as a function of $\xi$.}
      %evaluated from the eigenvalues $\varepsilon_k$ of $\mathcal{H}$ and those evaluated from the eigenvalues $-\pi \lambda_k$ of $\mathcal{H}_{BW}$
    \label{fig:deltasmallgrad}
\end{figure}
%%%%%%%%%%%%%%%%%%%%%%%%%%%%%%%%%%%%%%%%%%%%

To check the validity of our ansatz \eqref{eq:Ttilde}, in Fig.~\ref{fig:smallgrad} we compare both eigenvalues and eigenvectors of the scaled matrix $-\pi T$ to those of $H$,
for a fixed ratio $R=1/2$ and increasing $L$. On the left,
the difference of the low-lying spectra shows a similar pattern as for the $L\to\infty$ case in Fig.~\ref{fig:spectragrad}.
Plotting also the corresponding eigenvectors, one observes that the differences are almost invisible. We thus quantify them by calculating the deviation of their overlap from unity, as shown on the right of Fig.~\ref{fig:smallgrad}. In the relevant low-lying part of the spectrum they show a flat profile, with a very small error which does not increase with the size $L$. However, the discrepancies tend to grow towards the spectral edges, developing a maximum (not shown) for the largest negative eigenvalue. Nevertheless, the matrix elements of the commutator $[H,T]$ remain very small, as observed in our numerics.

The results for other values of the ratio turn out to be very similar. To illustrate this, we plot the entropy difference $\delta S$ in Fig.~\ref{fig:deltasmallgrad} for various $R$ and increasing $\xi$. The deviations are rather small and decay in an
oscillatory fashion. This further confirms that the discretized CFT ansatz provides a very good description of the lattice EH, even in the absence of an exact commutation property.

\subsection{Continuum limit}

To conclude this section, we provide an alternative check of the CFT result via a continuum limit, which can be understood intuitively. For a systematic derivation, we refer to Appendix \ref{app:CLgradient}.
As discussed earlier, the exact expression of the matrix $H$ in the EH \eqref{eq:EHff} contains, in general, long-range hopping terms. If the matrix elements $H_{i,i+r}$ were independent of $i$, these amplitudes would modify the dispersion
relation, leading to a renormalized Fermi velocity. Instead, if $H_{i,i+r}$ are assumed to vary slowly, one can invoke again the LDA argument, and calculate a space-dependent Fermi velocity. Finally, for the massless Dirac fermion, this velocity can simply be identified with the weight in \eqref{eq:EHhom} such that one has \cite{ETP-19}
\begin{equation}
\label{eq:CLbeta}
    2\pi \beta(x) =- 2a \sum_{r>0} r \sin(q_F a r) H_{i,i+r},
\end{equation}
where $x=ia$ and $a$ is the lattice spacing. The continuum limit can then be carried out by sending $a \to 0$, and keeping the length $aL$ fixed. The function $\beta(x)$ is thus not simply given by the nearest-neighbour hopping $H_{i,i+1}$, but rather as a weighted sum along the diagonals of the matrix $H$.

The generalization of the above procedure to inhomogeneous cases is straightforward \cite{RSC-22}. Indeed, one simply has to replace the constant Fermi momentum with the function $q_F(x)a=\arccos(x/\xi)$ obtained by LDA. Furthermore, as pointed out in section \ref{sec:CFT}, the resulting weight factor $\beta(x)=\tilde \beta(x) v_F(x)$ is now a product of the physical Fermi velocity and the local inverse temperature, such that the latter can be written as
\begin{equation}
\label{eq:CLbeta2}
    2\pi \tilde{\beta}(x) = -\frac{2 a}{v_F(x)}\sum_{r>0} r \sin(q_F(x)a r)H_{i,i+r}
\end{equation}
Finally, instead of the summation along the diagonals, it is numerically easier to consider a row-wise sum, which leads to the same result in the continuum limit. Fixing the length scale as
$La=1$ and using the continuous variable $x=(i-1/2)a$ one obtains
\begin{equation}
\label{eq:CLbeta3}
    \tilde{\beta}(x) = -\frac{1}{2\pi v_F(x)}\sum_{j\in A} (j-i) \sin[q_F(x)a(j-i)] \frac{H_{i,j}}{L}.
\end{equation}

\begin{figure}
\centering    \includegraphics[width=0.6\textwidth]{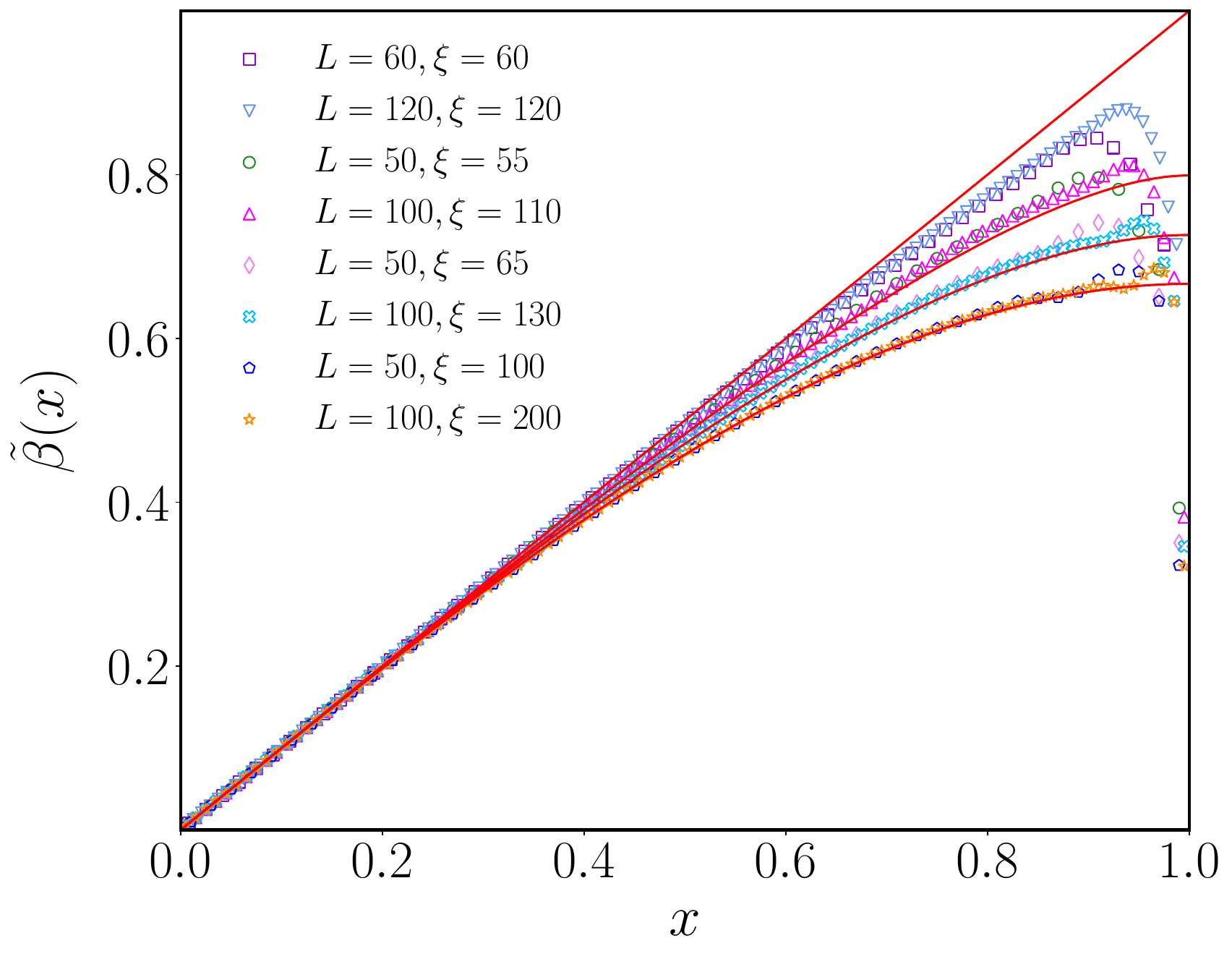}
      \caption{Spatial profile of the inverse temperature $\tilde{\beta}(x)$ plotted as a function of $x=(i-1/2)/L$. The numerical data (symbols) are obtained from the continuum limit \eqref{eq:CLbeta3} for different values of $L$ and $\xi$, and plotted against the CFT result (solid red line) in \eqref{eq:betagrad}.}
    \label{fig:CLimit}
\end{figure}

The continuum limit outlined above was studied for the equivalent domain-wall melting problem, finding perfect agreement with CFT in the regime $L \gg \xi$ \cite{RSC-22}. Here we extend these calculations to small gradients with ratio $R\le 1$, focusing on $s=x_0=0$. The results are shown in Fig.~\ref{fig:CLimit}, where the symbols represent the numerical data obtained by performing the sum in \eqref{eq:CLbeta3} along the rows of $H$, plotted as a function of $x=(i-1/2)/L$. We consider four different ratios $R=L/\xi$, and report two sets of data for each of them. These are compared to the CFT result \eqref{eq:betagrad} by substituting $L \to La=1$ and $\xi \to \xi a=1/R$, as shown by the red lines. In general, the agreement is good away from the boundary $x \approx 1$, where the lattice effects are still clearly visible. We also observe that the agreement improves for decreasing values of the ratio $R$, and the profile moves toward the sine function found in the limit of a finite homogeneous system $R \to 0 $ \cite{CT-16}. In contrast, the BW limit $R \to 1$ is approached very slowly as $L$ is increased, since the edge of the density profile coincides with the boundary of the chain, and lattice effects play an important role.

%%%%%%%%%%%%%%%%%%%%%%%%%%%%%%%%%%%%%%%%%%%%
%%%%%%%%%%%%%%%%%%%%%%%%%%%%%%%%%%%%%%%%%%%%

\section{Entanglement Hamiltonian for the harmonic trap\label{sec:trap}}

Let us now move to the study of the EH for the harmonic trap, focusing only on the thermodynamic limit $L \to \infty$. As discussed in Section \ref{sec:models}, the Fermi velocity \eqref{eq:vFtrap} differs only by an extra scale factor $\xi=\sqrt{2N}$ from that of the gradient chain. Since the scale enters the expression \eqref{eq:betainhom} of the inverse temperature only via the prefactor $\tilde L$, which is defined in \eqref{eq:tildeL} and depends on the inverse of $v_F(x)$,
one finds immediately
\begin{equation}
    \tilde \beta(x)=\frac{x-x_0}{\xi\sqrt{1- \left(\frac{x_0}{\xi} \right)^2}}=\frac{x-x_0}{v_F(x_0)}.
\end{equation}
Note that, analogously to the gradient chain, one has to set $L=\xi$ in the CFT formula to obtain the proper limit. In turn, the EH is given again by the BW form in \eqref{eq:EHLggxi}, only the expression of $v_F(x_0)$ changes.

Calculating the exact EH for the harmonic trap amounts to solve the eigenvalue problem of the integral operator $\hat{\mathcal{K}}$ in \eqref{eq:Kint}. Remarkably, in complete analogy to the homogeneous case \cite{SP-61,EP-13}, one finds again a second-order differential operator $\hat{D}$ that commutes with the integral operator, $[\hat{\mathcal{K}},\hat{D}]=0$, and thus has the same set of eigenfunctions. Such a commutation property was proven for integral operators with a kernel composed of classical orthogonal polynomials \cite{GRUNBAUM-83}. In particular, our kernel \eqref{eq:Kxy} corresponds to the Hermite case in \cite{GRUNBAUM-83}, after a proper symmetrization. The derivation of the commuting operator $\hat{D}$ is summarized in Appendix \ref{App:Dop}, and one finds
\begin{equation}
    \hat D = \frac{\dd}{\dd x}(x-x_0)\frac{\dd}{\dd x} - x^2(x-x_0) + 2N(x-x_0).
    \label{eq:Dtrap}
\end{equation}
Thus we observe that, up to a factor $(-1/2)$, it has exactly the same form as the Hamiltonian \eqref{eq:Htrap}, multiplied by a linear BW weight factor. Note that it is inserted between the derivatives in the first term, and one identifies $\mu=N$ in the last term. The BW result \eqref{eq:EHLggxi} can thus be formulated as
\begin{equation}
\label{eq:HharD}
    \mathcal{\hat H}_{BW} = -\frac{\pi}{v_F(x_0)}\hat D.
\end{equation}

In order to verify the validity of \eqref{eq:HharD}, one needs to determine the eigenvalues $\chi_k$ of $\hat{D}$ and compare it to those $\varepsilon_k$ of the exact EH in \eqref{eq:EHtrap}. While the latter can be easily obtained by diagonalizing the overlap matrix \eqref{eq:overlapHar}, a similar duality can be exploited also for the differential operator \cite{GRUNBAUM-83}. Namely, there exists a symmetric tridiagonal matrix $M$ whose spectrum is identical to that of $\hat{D}$. The expression of $M$ can be obtained by evaluating the matrix elements of $\hat{D}$ in the basis of the occupied oscillator eigenfunctions \eqref{eq:Phitrap}. As shown in Appendix \ref{app:dualM}, this yields
\begin{equation}
    M_{m,n} = 2(N-n)\sqrt{\frac{n}{2}} \delta_{m,n-1} +2(N-m)\sqrt{\frac{m}{2}} \delta_{n,m-1} - 2 \left(N-n-\frac{1}{2}\right) x_0 \delta_{m,n},
    \label{eq:Mmn}
\end{equation}
where $0 \le m,n < N$. Note that $M$ also commutes with the overlap matrix, $[\mathbb{A}, M]=0$, which is dual to the commutation relation between $\hat{\mathcal{K}}$ and $\hat{D}$ \cite{GRUNBAUM-83}. 

%%%%%%%%%%%%%%%%%%%%%%%%%%%%%%%%%%%%%%%%%
\begin{figure}
  \centering
    \hspace*{-0.25cm}
    \subfloat{\includegraphics[width=0.45\textwidth]{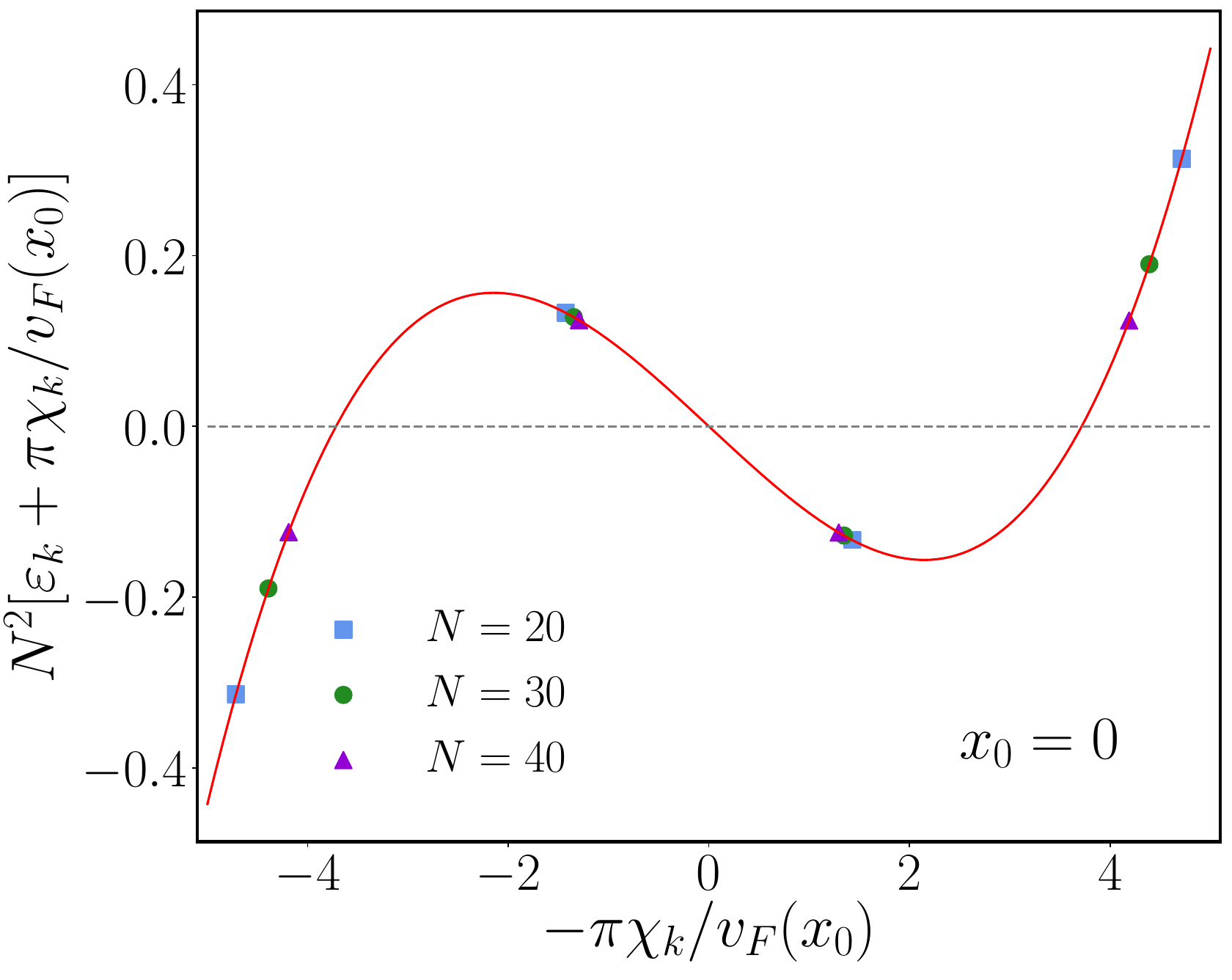}}    
    \hspace{0.5cm}\subfloat{\includegraphics[width=0.45\textwidth]{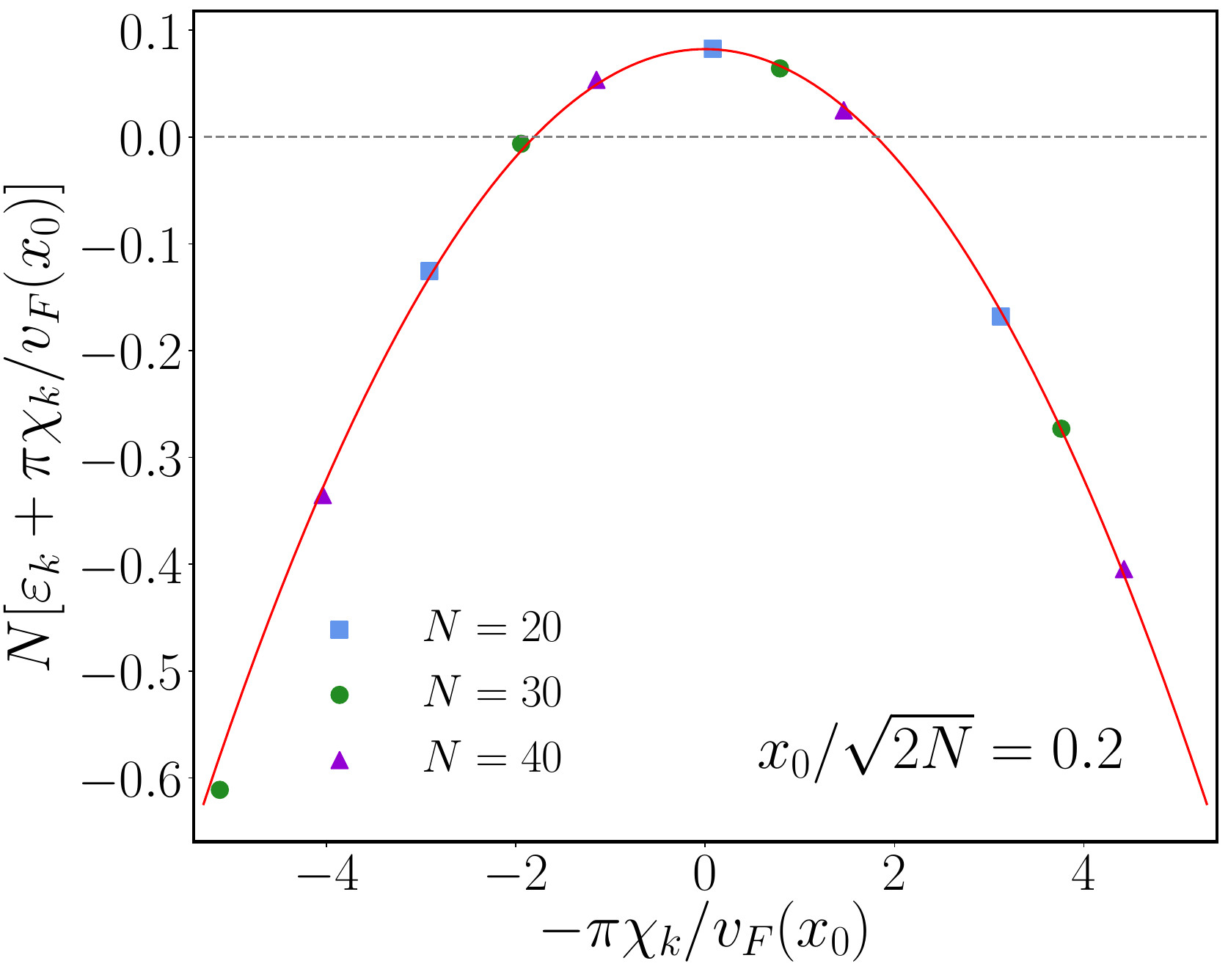}} 
   \caption{Rescaled differences between the spectra $\varepsilon_k$ and $-\pi\chi_k/v_F(x_0)$, plotted as function of $-\pi\chi_k/v_F(x_0)$ for $x_0=0$ (left) and $x_0/\xi=0.2$ (right). The symbols represent the numerical data obtained at different values of $N$, and the solid red lines are polynomial fits as described in the text.}
     \label{fig:scalspec}
\end{figure}
%%%%%%%%%%%%%%%%%%%%%%%%%%%%%%%%%%%%%%%%%

Calculating the eigenvalues $\chi_k$ of the matrix $M$, we observe that \eqref{eq:HharD} gives an excellent approximation of the low-lying entanglement spectrum $\varepsilon_k$ already for rather small particle number $N$.
%agreement between the spectra $\varepsilon_k$ and $$ in the low-energy region.
By analogy with the homogeneous case \cite{E-23}, we expect that the corrections to the BW result can be written in terms of a series expansion
\begin{equation}
    \mathcal{\hat H}=\mathcal{\hat H}_{BW} + \sum_{p=1}^{\infty}\frac{1}{N^p} P_{p+1}(\mathcal{\hat H}_{BW}),
    \label{eq:EHtrapcorr}
\end{equation}
where $P_p(x)$ denotes an $p$-th order polynomial that depends on $x_0$. Even though their analytical form is not known, we can extract the lowest order correction numerically, as demonstrated in Fig. \ref{fig:scalspec}. In particular, we find that for the symmetric bipartition, $x_0=0$, the first-order correction vanishes identically, $P_2(x)\equiv 0$, which is due to the particle-hole symmetry of the spectrum. The leading corrections to the BW spectra are then shown in the left panel of Fig. \ref{fig:scalspec}, rescaled by $N^2$ and plotted as a function of $x=-\pi \chi_k/v_F(x_0)$. While the symbols correspond to the numerical data for different $N$, the red curve is obtained by fitting a third order polynomial to the set $N=40$, with the result $P_3(x)=-0.1093 x + 0.0079 x^3$. On the right of Fig. \ref{fig:scalspec} we show the corrections for a nonsymmetric bipartition, keeping the ratio $x_0/\xi$ fixed for different $N$.
We observe that in this case the leading correction is $\mathcal{O}(N^{-1})$, as demonstrated by the rescaled data, and the fitted polynomial is $P_2(x)=0.0636-0.0196 x^2$.

\begin{figure}
  \centering
    \hspace*{-0.25cm}
    \subfloat{\includegraphics[width=0.45\textwidth]{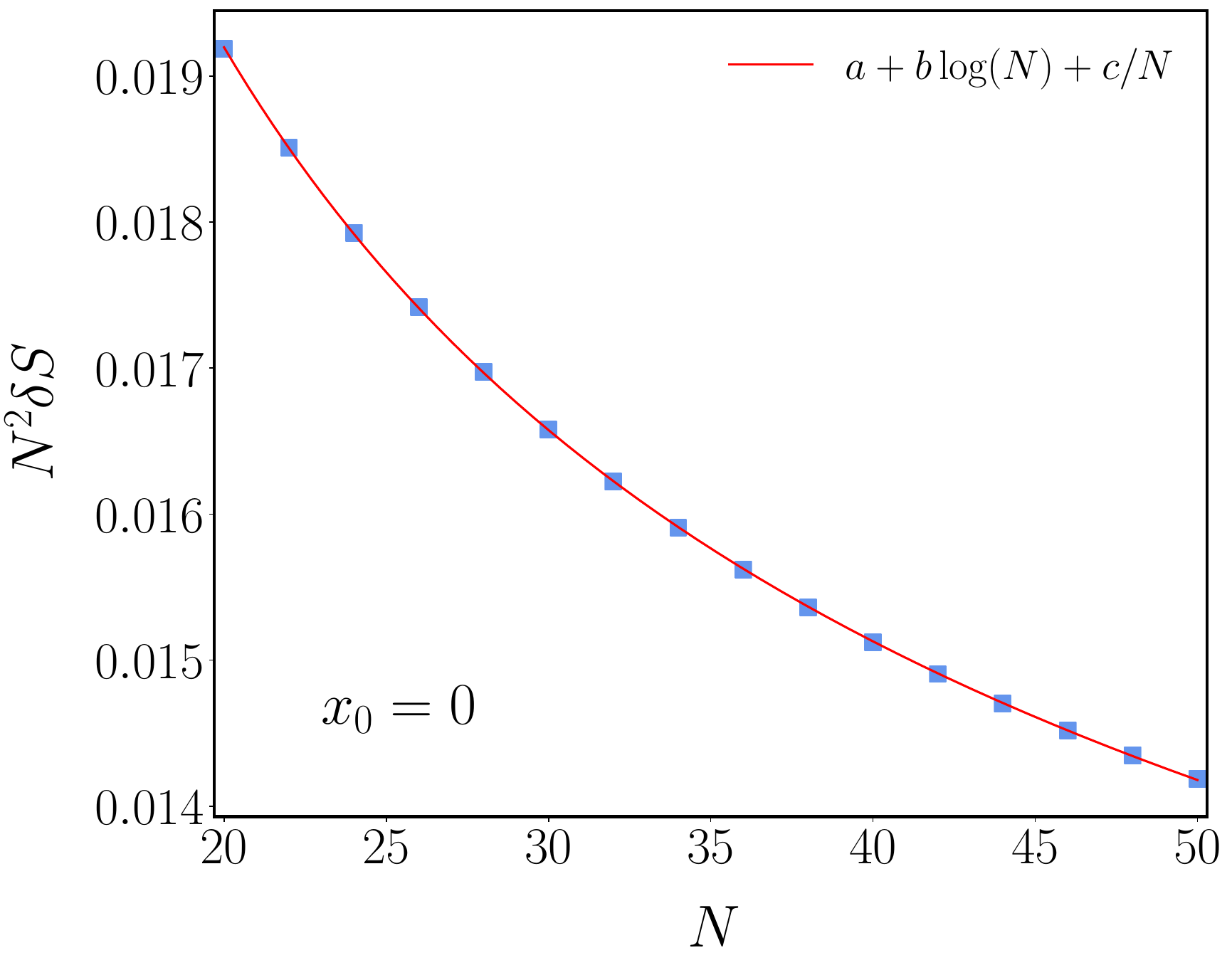}}    \hspace{0.5cm}
    \subfloat{\includegraphics[width=0.45\textwidth]{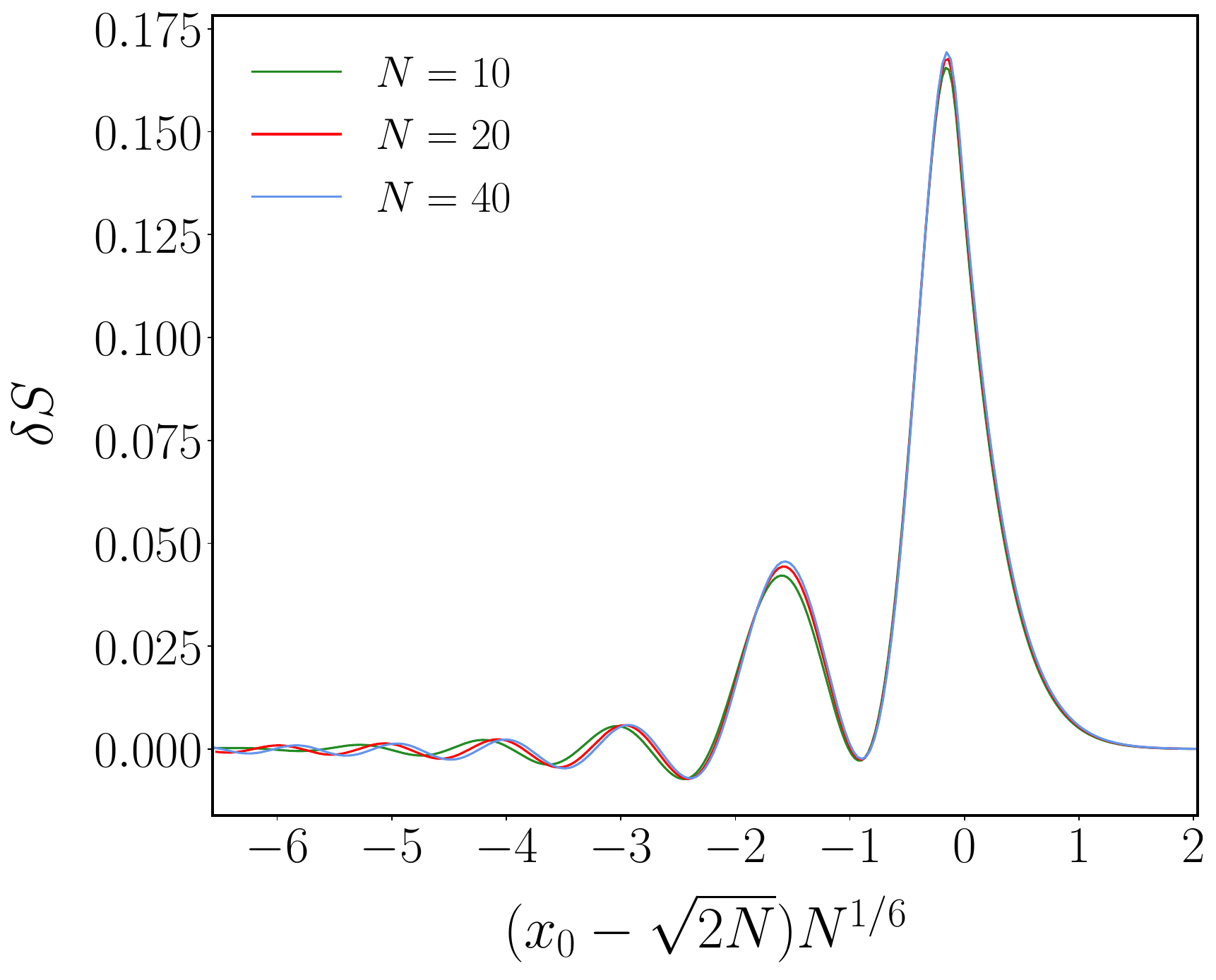}} 
   \caption{\textit{Left:} Scaled entropy deviation $N^2\delta S$ for $x_0=0$ as function of $N$. The solid red line is a fit to the ansatz \eqref{eq:dStrap}, with parameters $a=0.018$, $b=-0.0016$, and $c=0.1181$.  \textit{Right:} Entropy difference as function of the edge scaling variable $(x_0-\sqrt{2N})N^{1/6}$ for increasing values of $N$.}
     \label{fig:dSHar}
\end{figure}

The BW prediction \eqref{eq:HharD} can be further tested by studying the deviation of the entropy. Similarly to the gradient chain, $S$ is obtained from \eqref{eq:S} while the approximation $S_{BW}$ follows by replacing $\varepsilon_k \to -\pi \chi_k/v_F(x)$. Their difference $\delta S$ for $x_0=0$ is found to be extremely small and decreasing as a function of $N$, and we put forward the ansatz
\begin{equation}
    \delta S = \frac{a + b \ln(N)}{N^2} + \mathcal{O}(N^{-3}),
    \label{eq:dStrap}
\end{equation}
suggested by \eqref{eq:EHtrapcorr} with vanishing $P_2(x)= 0$.
% , and we find a very good agreement for their values in the bulk.
On the left of Fig. \ref{fig:dSHar}, the symbols correspond to the numerical entropy deviation $\delta S$ rescaled by $N^2$, and the line shows a fit to the ansatz \eqref{eq:dStrap}. The rescaled data is almost constant ($a\approx 0.018$), with a slight decrease due to a small negative logarithmic term ($b\approx -0.0016$) and a subleading correction. Note that the fit clearly gets worse when excluding the logarithm. Its presence can be associated to the scaling $\rho(\varepsilon)\propto \ln (N)/(2\pi^2)$ of the density of states in the entanglement spectrum, which yields the leading logarithmic contribution to the bulk entropy \cite{Vicari-12,CLDM-15,DSVC-17}. As expected, $\delta S$ remains very small in the entire bulk region, with significant deviations only in the edge regime $x_0 \to \xi$. Here one can repeat the analysis of the previous section, with the result shown on the right of Fig. \ref{fig:dSHar}. Plotted as a function of the proper edge scaling variable $(x_0-\sqrt{2N})N^{1/6}$ \cite{TW-94,LLDMS-18,DLDMS-19}, the curves show an almost perfect collapse already for moderate values of $N$. In fact, it seems to be the continuous version of Fig.~\ref{fig:deltaSedge}, strongly suggesting that the anomalous behaviour of the spectra in the edge regime may have some universal description, that goes beyond CFT.

%%%%%%%%%%%%%%%%%%%%%%%%%%%%%%%%%%%%%%%%%
%%%%%%%%%%%%%%%%%%%%%%%%%%%%%%%%%%%%%%%%%

\section{Discussion\label{sec:disc}}

We studied the entanglement Hamiltonian for free-fermion systems with slowly varying potentials, which induce an inhomogeneous density. In particular, we considered a hopping chain with a potential gradient and a continuous Fermi gas in a harmonic trap, which turn out to behave very similarly. Indeed, they admit an effective curved-space CFT description with a space-dependent Fermi velocity, which yields the EH in a Bisognano-Wichmann form for both cases. Although both the lattice and the continuum model breaks the relativistic symmetry, the BW structure is inherited by operators that commute exactly with the EH. Furthermore, after an appropriate rescaling with the value of the Fermi velocity at the subsystem boundary, they reproduce the low-lying entanglement spectra and entropies very accurately. The approximation only breaks down if the entanglement cut is located within the dilute edge regime of the density profile, where the LDA-based CFT treatment is not applicable. Nevertheless, our numerical results suggest some universality in the deviations from the BW ansatz, which deserves further investigation.

For the gradient chain we also explored finite-size effects, i.e. the case where the length scale of the density profile exceeds the chain size. Constructing a tridiagonal matrix according to the CFT prescription, we observe that the matrix elements of its commutator with the EH remain very small. This raises the question whether an exactly commuting operator exists even in the presence of a boundary. This is indeed the case in the limit of a homogeneous open chain ($\xi \to \infty$), where the tridiagonal matrix has the CFT expression \eqref{eq:betahom}, with the substitution $2L \to 2L+1$ \cite{EP-18}. This example demonstrates the subtleties involved with the discretization of the CFT result.

Another interesting feature of the gradient chain is the duality with the EH of a disk in a two-dimensional \emph{homogeneous} Fermi gas \cite{E-23}. Indeed, the entanglement spectrum of the subsystem $[s+1,\infty)$ is identical to that of the 2D system in the sector of angular momentum $s$, while the length scale $\xi=q_F R$ plays the role of the disk radius multiplied by the Fermi wavenumber. This can be proved by a direct comparison of the underlying (discrete and continuous) Bessel kernels \cite{MNS-19}. Furthermore, the continuous Bessel kernel commutes with a differential operator $\hat D_s$ \cite{Slepian-64}, and our numerical study of its eigenvalues using \cite{Lederman-17} suggests that it is precisely dual to the $T$ matrix with elements in \eqref{eq:Telements}. The origin of this remarkable duality remains to be understood, and its analytic derivation to be provided.

It would also be useful to extend these studies to more general potentials. While an exact commutation property is unlikely to be found for the generic case, the CFT prediction \eqref{eq:betainhom} for the inverse temperature could still provide an accurate ansatz for the EH. Finally, one could explore the effect of interactions for the gradient chain, which can be effectively described by an inhomogeneous Luttinger liquid \cite{EB-17,BLDS-20}. However, the question how the spatial variation of the Luttinger parameter can be implemented in the CFT machinery for the EH is still open.

\begin{acknowledgments}

We thank P. A. Bernard, G. Parez, I. Peschel, G. Sierra and E. Tonni for fruitful discussions. The authors acknowledge funding from the Austrian Science Fund (FWF) through project No. P35434-N.

\end{acknowledgments}

\appendix

\section{Continuum limit for the gradient chain}
\label{app:CLgradient}

In this appendix, we provide the derivation of the continuum limit of the EH for the gradient chain. Following \cite{ETP-19}, we set $d(i)=H_{i,i}$ and $H_{i,i+r}=t_r(i+r/2)$ for $r>1$, where $i+r/2$ is the midpoint between sites $i$ and $i+r$, and rewrite the EH in \eqref{eq:EHff} as
\begin{equation}
\label{eq:HCLapp}
    \mathcal{H}= \sum_i d(i)c_i^{\dagger}c_i+\sum_i \sum_{r \geq 1} t_r(i+r/2)\left(c_i^{\dagger}c_{i+r}+c_{i+r}^{\dagger}c_i \right).
\end{equation}
 In order to replace the lattice variable with a continuous one $a(i-1/2)\rightarrow x$, we rewrite the fermionic operators in terms of left and right moving fields $\psi_{L,R}(x)$,  
 \begin{equation}
 \begin{split}
 \label{eq:Clcop}
 c_i &\rightarrow  \sqrt{a}\left( \eE^{\ir \varphi(x) }\psi_R(x)+\eE^{-\ir \varphi(x) }\psi_L(x)\right)\\
  c_{i+r} &\rightarrow  \sqrt{a}\left( \eE^{\ir\varphi(x+ra)}\psi_R(x+ra)+\eE^{-\ir \varphi(x+ra)}\psi_L(x+ra)\right),
 \end{split}
 \end{equation}
where $\varphi(x)$ is an integrated phase \cite{RSC-22},
\begin{equation}
    \varphi(x)=\int^x \dd x q_F(x).
\end{equation}
Since in \eqref{eq:HCLapp} only the phase differences enter, and the Fermi momentum is a slowly varying function of the position, we can approximate
\begin{equation}
    \varphi(x+ra)-\varphi(x) \approx ra q_F(x+ra/2).
\end{equation}
In taking the continuum limit, the on-site amplitude goes to $d(i)\rightarrow d(x)$, and  the hopping amplitudes can be expanded as $t(i+r/2)\rightarrow t(x)+ar t'(x)/2+ \mathcal{O}(a^2)$. Inserting the above approximations in \eqref{eq:HCLapp}, we obtain the following expression
\begin{equation}
   \begin{split}
        \mathcal{H}=\int &\dd x d(x) [\psi^{\dagger}_R(x)\psi_R(x)+\psi_L^{\dagger}(x)\psi_L(x)]+\sum_{r=1}^{\infty}\int \dd x  \left( t_r(x)+\frac{ra}{2}t_r'(x)\right)\\
        \times \Big\{  & \cos(r a q_F(x+\frac{ra}{2}))   [2(\psi^{\dagger}_R(x)\psi_R(x)+\psi_L^{\dagger}(x)\psi_L(x))]+ r a \partial_x (\psi^{\dagger}_R(x)\psi_R(x)+\psi_L^{\dagger}(x)\psi_L(x))  + \\
         +& \sin(r a q_F(x+\frac{ra}{2}))ra [ \ir (   \psi^{\dagger}_R(x)\psi'_R(x)+\psi_L^{\dagger}(x)\psi'_L(x))+ \mathrm{h.c.} ]\Big\}.
   \end{split}
\end{equation}
Expanding the cosine and sine functions, we obtain
\begin{equation}
    \cos(raq_F(x+ra/2))\approx \cos(ra q_F(x))- \frac{ra}{2}q_F'(x)ar \sin(ra q_F(x)), \quad \sin(raq_F(x+ra/2))\approx \sin(raq_F(x)) 
\end{equation}
and keeping only the leading order term in $a$, the expression above becomes
\begin{equation}
    \begin{split}
    \label{eq:HCLall}
        \mathcal{H}=&\int \dd x d(x) [\psi^{\dagger}_R(x)\psi_R(x)+\psi_L^{\dagger}(x)\psi_L(x)]\\
        + \sum_{r=1}^{\infty} \Big[ & \int \dd x  \left( t_r(x)+\frac{ra}{2}t_r'(x)\right) \cos(ra q_F(x))2 (\psi_R^{\dagger}(x)\psi_R(x)+\psi_L^{\dagger}(x)\psi_L(x))\\
        +&\int \dd x  t_r(x)\cos(ra q_F(x))ra\partial_x(\psi^{\dagger}_R(x)\psi_R(x)+\psi_L^{\dagger}(x)\psi_L(x))\\
        -&\int \dd x t_r(x)\frac{ra}{2}q_F'(x)ra\sin(raq_F(x))2(\psi^{\dagger}_R(x)\psi_R(x)+\psi_L^{\dagger}(x)\psi_L(x))\\
        -&\int \dd x  t_r(x) a r \sin(ra q_F(x))  T_{00}(x)\Big]+ \mathcal{O}(a^2),
    \end{split}
\end{equation}
where we replaced the expression of the energy density operator,
\begin{equation}
    \label{eq:T00}
    T_{00}(x)=\frac{1}{2}[\psi_R^{\dagger}(x)(-\ir \partial_x)\psi_R(x)-\psi_L^{\dagger}(x)(-\ir \partial_x)\psi_L(x)+\mathrm{h.c.}].
\end{equation}
We observe that three of the terms in \eqref{eq:HCLall} can be obtained as total derivative of the function
\begin{equation}
\sum_{r=1}^{\infty}t_r(x)\cos(r a q_F(x))(\psi_R^{\dagger}(x)\psi_R(x)+\psi_L^{\dagger}(x)\psi_L(x)),
\end{equation}
which then gives a vanishing boundary term if $t_r(x_0)=0$. The remaining terms can be written as
\begin{equation}
\label{eq:HCLintegral}
    \mathcal{H}=\int \dd x \mu(x)[\psi_R^{\dagger}(x)\psi_R(x)+\psi_L^{\dagger}(x)\psi_L(x)]+\int \dd x v(x)T_{00}(x), 
\end{equation}
with chemical potential and velocity defined as
\begin{equation}
\label{eq:mu}
    \mu(x)=d(x)+2\sum_{r=1}^{\infty} \cos(ra q_F(x)) t_r(x), \hspace{1cm} v(x)=-2 a\sum_{r=1}^{\infty}  r\sin(ra q_F(x)) t_r(x).
\end{equation}
One can now identify $v(x)=2 \pi \beta(x)$, as in \eqref{eq:CLbeta} of the main text. 
We note that the term $\mu(x)$ can also be evaluated numerically, and it is found to go to zero for increasing $\xi$.

Finally we illustrate the continuum limit on the example of the operator built from the commuting $T$ matrix \eqref{eq:Telements}
\begin{equation}
\label{eq:Top}
\mathcal{T}= \sum_{i,j=1}^L T_{i,j}c_i^{\dagger}c_j,
\end{equation}
which describes a hopping model similar to that of the physical Hamiltonian \eqref{eq:Hgrad}, but with hopping and diagonal terms multiplied by a factor increasing linearly from the boundary. The $\mathcal{T}$ operator commutes with $\mathcal{H}$ in \eqref{eq:EHff} by construction, since $[T,H]=0$.
Using the expression of the  matrix elements \eqref{eq:Telements}, we can identify $d_i=d(i)$ and $t_i=t(i+1/2)$, whereas the hopping for $r>1$ is identically zero. Taking the continuum limit, one then has
\begin{equation}
\label{eq:dt1}
   d(x)=-\frac{2x(x-x_0)}{\xi a}, \quad \quad  t(x)= \frac{x-x_0}{a}
\end{equation}
and the expressions in \eqref{eq:mu} simplify to
\begin{equation}
    \mu(x)=d(x)+2 \cos(a q_F(x)) t(x), \hspace{1cm} v(x)=-2 a \sin(a q_F(x)) t(x).
\end{equation}
Sustituting \eqref{eq:dt1} and using the relations $\cos(a q_F(x))=x/\xi$ and $\sin(a q_F(x))=v_F(x)$, we have $\mu(x)=0$ and 
\begin{equation}
\label{eq:TCL}
    \mathcal{T}=-2\int \dd x  (x-x_0)v_F(x)T_{00}(x).
\end{equation}
 Comparing this expression with \eqref{eq:HBW} we find that 
\begin{equation}
\mathcal{H}_{BW} = -\frac{\pi}{v_F(x_0)}\mathcal{T}.
\end{equation}
This is just the same relation found for the low-lying spectra of $H$ and $T$, and thus the continuum limit of $\mathcal{T}$ gives immediately the BW form of $\mathcal{H}$.

% Thus, in agreement with the choice of the origin of $A$ in the midpoint between the sites of the two subsystems, we choose the diagonal elements to be $d_i=-2/\xi(i-1/2)(i-1/2+s)$, so the constant shift in \eqref{eq:Telements} is equal to $2/\xi(s/2-1/4)$. The key observation is that, due to the presence of linear factors entering the elements of the tridiagonal marix, the structure of the $\mathcal{T}$ operator is the same as the EH in \eqref{eq:EHLggxi}.  In analogy to the homogeneous case, this leads us to conjecture the following relation between the eigenvalues $\varepsilon_k$ of \eqref{eq:Hhopp} and those of $T$ denoted as $\lambda_k$,

\section{Commuting differential operator for the harmonic trap}
\label{App:Dop}

Here, we recapitulate the result of Ref. \cite{GRUNBAUM-83}, proving the existence of a commuting differential operator associated to a certain integral operator, whose kernel is built from orthogonal polynomials. For simplicity, we shall focus on the case of Hermite polynomials. Although the problem in \cite{GRUNBAUM-83} is formulated in terms of time- and band-limited signals, it can be immediately translated into our setting with the harmonic trap. Namely, the discrete time variable corresponds to the mode index $n$, while the continuous frequency plays the role of space $x$. Below we translate the results into these variables to make the connection with our setting immediate.

Let us consider the integral operator $\mathcal{\hat K}_0$, with its kernel defined via the Hermite polynomials as
\begin{equation}
    K_0(x,y) = \sum_{n=0}^{N-1} c_n^{-1} H_n(x)H_n(y) \eE^{-y^2},
\end{equation}
and acting on the domain $(x_0,\infty)$. Note that, in contrast to Ref. \cite{GRUNBAUM-83}, we include the Gaussian weight function into the kernel, instead of the integration measure. The corresponding integral operator commutes with the second-order differential operator \cite{GRUNBAUM-83}
\begin{equation}
    \hat D_0 = \eE^{x^2}\frac{\dd}{\dd x}\left[(x-x_0) \eE^{-x^2}\frac{\dd}{\dd x}\right] + 2(N-1)x \, .
    \label{eq:D0}
\end{equation}
However, a comparison with \eqref{eq:Kxy} shows, that $K_0(x,y)$ is not quite the correlation kernel we need, as the Gaussian factors do not appear in a symmetric way. Nevertheless, one can easily show that the integral operator with the symmetric kernel follows from a simple similarity transformation
\begin{equation}
    \mathcal{\hat K} = \hat A \, \mathcal{\hat K}_0 \hat A^{-1} , \qquad
    \hat A \, f(x) = \eE^{-\frac{x^2}{2}} f(x).    
\end{equation}
Defining a new differential operator $\hat D=\hat A \, \hat D_0 \hat A^{-1}+C$ by the same transformation and an additive constant $C$, the commutation relation $[\mathcal{\hat K}, \hat D]=0$ follows immediately from $[\mathcal{\hat K}_0, \hat D_0]=0$. The similarity transformation yields
\begin{equation}
    \hat A \, \hat D_0 \hat A^{-1}=\eE^{\frac{x^2}{2}} \frac{\dd}{\dd x}\left((x-x_0) \eE^{-x^2}\frac{\dd}{\dd x}\right)\eE^{\frac{x^2}{2}} + 2(N-1)x,
\end{equation}
and using the commutation property
\begin{equation}
    \frac{\dd}{\dd x} \eE^{\pm\frac{x^2}{2}} = \eE^{\pm\frac{x^2}{2}} \left(\frac{\dd}{\dd x}\pm x\right)    
\end{equation}
one arrives at
\begin{equation}
    \hat A \, \hat D_0 \hat A^{-1}=
\left(\frac{\dd}{\dd x}-x\right)(x-x_0) \left(\frac{\dd}{\dd x}+x\right) + 2(N-1)x \, .
\end{equation}
It is easy to see that, with the choice $C=-(2N-1)x_0$, the operator $\hat D$ is given by \eqref{eq:Dtrap} reported in the main text.

It remains to prove the identity $[\mathcal{\hat K}_0, \hat D_0]=0$. Let us first write the action of the operators explicitly
\begin{align}
&\hat D_0 \mathcal{\hat K}_0 f(x) = 
\int_{x_0}^{\infty} \dd y \, D_0(x) \, K_0(x,y) \, f(y) \\
&\mathcal{\hat K}_0 \hat D_0 f(x) = 
\int_{x_0}^{\infty} \dd y \, K_0(x,y) D_0(y) f(y),
\end{align}
for an arbitrary function $f$. The notation $D_0(y)$ indicates that the operator \eqref{eq:D0} must be applied with variables corresponding to the argument. In the second line, integrating by parts twice one obtains
\begin{align}
\mathcal{\hat K}_0 \hat D_0 f(x) = 
&\int_{x_0}^{\infty} \dd y \left[-\frac{\dd}{\dd y}\left( K_0(x,y)\eE^{y^2}\right)(y-x_0)\eE^{-y^2}
\frac{\dd}{\dd y} + 2(N-1)y K_0(x,y)\right]f(y)= \nonumber \\
&\int_{x_0}^{\infty} \dd y \left[ \frac{\dd}{\dd y}\left((y-x_0)\eE^{-y^2}\frac{\dd}{\dd y} K_0(x,y)\eE^{y^2}\right)
+ 2(N-1)y K_0(x,y) \right]f(y)= \nonumber \\
&\int_{x_0}^{\infty} \dd y \, \eE^{-y^2} D_0(y) \left[K_0(x,y)\eE^{y^2} \right] f(y) \, .
\end{align}
Note that we used that both of the boundary terms vanish
\begin{equation}
    \left. (y-x_0) K_0(x,y)\frac{\dd f(y)}{\dd y}\right|_{x_0}^{\infty}=
    \left. (y-x_0) \eE^{-y^2}\frac{\dd}{\dd y}\left( K_0(x,y)\eE^{y^2}\right)f(y)\right|_{x_0}^{\infty}=0
\end{equation}

In other words, for the commutator to vanish, one needs to show that
\begin{equation}
    D_0(x) \left[ K_0(x,y)\eE^{y^2}\right] =
    D_0(y) \left[K_0(x,y)\eE^{y^2}\right].
\end{equation}
This identity has been proven in \cite{GRUNBAUM-83} for generic orthogonal polynomials. Alternatively, it can be directly verified in the case of Hermite polynomials using the Christoffel-Darboux formula to evaluate the sum in $K_0$, that gives
\begin{equation}
K_0(x,y)\eE^{y^2}= \frac{H_N(x)H_{N-1}(y)-H_{N-1}(x)H_N(y)}{2 c_{N-1}(x-y)}
\end{equation}
and applying the differential operator in the two different variables on both sides of the equality. The proof then follows after simple algebra.

\section{Dual tridiagonal matrix for the harmonic trap}
\label{app:dualM}

In this appendix, we provide the derivation of the dual tridiagonal matrix $M$ for the harmonic trap. As pointed out in sec.~\ref{sec:trap} and Ref. \cite{GRUNBAUM-83}, it is obtained by evaluating the matrix elements of $\hat{D}$ in the basis of the wavefunctions \eqref{eq:Phitrap}
\begin{equation}
\label{eq:Mmat}
    M_{m,n} = \int_{-\infty}^{\infty} \dd x \, 
    \Phi_m(x) \hat D \, \Phi_n(x).
\end{equation}
Using the derivative rule of the Hermite polynomials
\begin{equation}
    \frac{\dd}{\dd x}H_n(x)=2n H_{n-1}(x),
    \label{eq:derHer}
\end{equation}
one obtains for the oscillator function
\begin{equation}
    \frac{\dd}{\dd x}\Phi_n(x) =
    -x \Phi_n(x)+2n \sqrt{\frac{c_{n-1}}{c_n}}\Phi_{n-1}(x)=
    -x \Phi_n(x)+\sqrt{2n} \Phi_{n-1}(x),
\end{equation}
where the first term comes from the derivative of the Guassian and we used $c_n=\sqrt{\pi}2^{n} n!$. 
We will also need the second derivative which can be simply obtained from the eigenvalue equation $\hat H \Phi_n = E_n \Phi_n$ of the Hamiltonian \eqref{eq:Htrap}. Using $E_n=n+1/2$, one finds
\begin{equation}
    \frac{\dd^2}{\dd x^2}\Phi_n(x)=(x^2-2n-1)\Phi_n(x).
\end{equation}
Hence, for the differential operator defined in \eqref{eq:Dtrap} we get
\begin{equation}
    \hat D \, \Phi_n\, (x)=
    -x \Phi_n(x)+2 \sqrt{\frac{n}{2}} \Phi_{n-1}(x)+2
    (N-n-1/2)(x-x_0)\Phi_n(x).
\end{equation}
Using the property of the Hermite wavefunctions
%
%\begin{equation}
 %   \hat D \, \Phi_n\, (x)=
 %   \frac{x}{2} \Phi_n(x)-\sqrt{\frac{n}{2}} \Phi_{n-1}(x)-
 %   (N-n-1/2)(x-x_0)\Phi_n(x)
%\end{equation}
%
%
\begin{equation}
    \int_{-\infty}^{\infty}\dd x \, \Phi_m(x) x \, \Phi_n(x)=
    \sqrt{\frac{n}{2}} \delta_{m,n-1} + \sqrt{\frac{m}{2}} \delta_{n,m-1},
%\begin{cases}
%\sqrt{\frac{n+1}{2}}, \hspace{1 cm} \text{if} \quad m=n+1,\\
%\sqrt{\frac{n}{2}},\hspace{1 cm} \text{if} \quad m=n-1,\\
%0 ,\hspace{2 cm} \text{otherwise}
%\end{cases}
\end{equation}
as well as their orthogonality, one can finally evaluate the matrix elements in \eqref{eq:Mmat}, obtaining the expression 
\begin{equation}
    M_{m,n} = 2(N-n)\sqrt{\frac{n}{2}} \delta_{m,n-1} +2(N-m)\sqrt{\frac{m}{2}} \delta_{n,m-1} - 2 \left(N-n-\frac{1}{2}\right) x_0 \delta_{m,n}.
    \label{eq:Mmnapp}
\end{equation}
%

%\begin{equation}
%    M_{m,n} = -(N-n)\sqrt{\frac{n}{2}} \delta_{m,n-1} -(N-m)\sqrt{\frac{m}{2}} \delta_{n,m-1} + (N-n-1/2)x_0 \delta_{m,n}
%\end{equation}

From the matrix elements of $M$, we can explicitly prove the commutation relation with the overlap matrix $\mathbb{A}$. Since the two matrices are symmetric, proving their commutation property,  $[M,\mathbb{A}]=0$ is equivalent to prove that their product is symmetric, $(M \mathbb{A})_{m,n}=(M \mathbb{A})_{n,m}$. 
Since $M$ is tridiagonal, the only nonzero terms in the sum are
\begin{equation}
\label{eq:MA}
\begin{split}
(M\mathbb{A})_{m,n}=%&M_{m,m-1}\mathbb{A}_{m-1,n}+M_{m,m}\mathbb{A}_{m,n}+M_{m,m+1}\mathbb{A}_{m+1,n}\\
&2(N-m)\sqrt{\frac{m}{2}}\frac{1}{\sqrt{c_{m-1}c_n}}\int_{x_0}^{\infty}\eE^{-x^2}H_{m-1}(x)H_{n}(x)+\\
-&2\left(N-m-\frac{1}{2}\right)x_0 \frac{1}{\sqrt{c_{m}c_n}}\int_{x_0}^{\infty}\eE^{-x^2}H_{m}(x)H_{n}(x)+\\
+&2(N-m-1)\sqrt{\frac{m+1}{2}}\frac{1}{\sqrt{c_{m+1}c_n}}\int_{x_0}^{\infty}\eE^{-x^2}H_{m+1}(x)H_{n}(x).
\end{split}
\end{equation}
Using $c_{m-1}=c_m/2m$ and $c_{m+1}=2(m+1)c_m$, as well as the recursion relation for Hermite polynomials
\begin{equation}
    H_{n+1}(x)= 2x H_n(x) -2n H_{n-1}(x),
    \label{eq:recursion}
\end{equation}
we can rewrite the expression above as
\begin{equation}
    \begin{split}
      % =   \frac{1}{\sqrt{h_mh_n}}\int_{x_0}^{\infty} \dd x \eE^{-x^2}&[ 2(N-m)m H_{m-1}(x)H_n(x) - 2\left(N-m-\frac{1}{2}\right)x_0H_m(x)H_n(x)+\\
 % &+ (N-(m-1))H_{m+1}(x)H_n(x)]=\\
(M\mathbb{A})_{m,n}=  \frac{1}{\sqrt{c_mc_n}}\int_{x_0}^{\infty} \dd x \eE^{-x^2}[&-2\left(N-m-\frac{1}{2}\right)x_0H_m(x)H_n(x)\\&+(N-m-1)2xH_m(x)H_n(x)+H'_m(x)H_n(x)],
    \end{split}
\end{equation}
where we used also the property \eqref{eq:derHer}.
In order to prove that the expression above is symmetric, we subtract the one obtained by exchanging $m \leftrightarrow n$,
\begin{equation}
    \begin{split}
[M,\mathbb{A}] =\frac{1}{\sqrt{c_mc_n}}&\left\{  2(m-n)\int_{x_0}^{\infty} \dd x 
\eE^{-x^2}(x_0-x)H_m(x)H_n(x) \right.\\
&\left.+\int_{x_0}^{\infty} \dd x 
\eE^{-x^2}[H'_m(x)H_n(x)-H_n'(x)H_m(x)]   \right\}.
\end{split}
\label{eq:AMMA}
\end{equation}
In the first integral we can use again \eqref{eq:derHer} and subsequently take the derivative of the recursion relation \eqref{eq:recursion},
\begin{equation}
    2(m+1) H_m(x)=H'_{m+1}(x)=2xH_m'(x)+2H_m(x)-H_m''(x),
\end{equation}
so that it becomes
\begin{equation}
    \begin{split}
        &\frac{1}{\sqrt{c_mc_n}}\int_{x_0}^{\infty}\dd x (x_0-x)\eE^{-x^2} 2x[H'_m(x)H_n(x)-H'_n(x)H_m(x)] \\
        -&\frac{1}{\sqrt{c_mc_n}}\int_{x_0}^{\infty}\dd x (x_0-x)\eE^{-x^2}[H_m''(x)H_n(x)-H_n''(x)H_m(x)].
    \end{split}
\end{equation}
Observing that $2x \eE^{-x^2}=-\frac{\dd}{\dd x}\eE^{-x^2}$, we can integrate by parts in the first integral of the expression above, and obtain
\begin{equation}
    \begin{split}
        &-\frac{1}{\sqrt{c_mc_n}} \eE^{-x^2}(x_0-x)[H'_m(x)H_n(x)-H_n'(x)H_m(x)]  \Big|_{x_0}^{\infty}+\\
        &-\frac{1}{\sqrt{c_mc_n}}\int_{x_0}^{\infty}\dd x \eE^{-x^2}[H'_m(x)H_n(x)-H'_n(x)H_m(x)].
    \end{split}
\end{equation}
In this expression, the boundary terms vanish when evaluated at $x=x_0$ and $x=\infty$ and the integral exactly cancels the last term in \eqref{eq:AMMA}. This proves the commutation relation.

%
%\begin{equation}
 %   \begin{split}
%        [M,A]=&\frac{1}{\sqrt{c_mc_n}}\int_{x_0}^{\infty}\dd x \eE^{-x^2}(x_0-x)[2xH'_m(x)H_n(x)-2xH'_n(x)H_m(x)]+\\
%        &-\frac{1}{\sqrt{c_mc_n}}\int_{x_0}^{\infty}\dd x \eE^{-x^2}(x_0-x)[H_m''(x)H_n(x)-H_n''(x)H_m(x)]+ \\ &+\frac{1}{\sqrt{c_mc_n}}\int_{x_0}^{\infty}\dd x \eE^{-x^2}[H'_m(x)H_n(x)-H'_n(x)H_m(x)].
%    \end{split}
%\end{equation}
%

%\bibliographystyle{unsrt.bst}
%\bibliographystyle{apsrev4-2}

%

\end{document}